\documentclass[
reprint, 
longbibliography,
 showpacs,
 amsmath,amssymb,
 aps,
 pre
floatfix,
]{revtex4-2}

\usepackage{amsmath,amssymb}
\usepackage{physics}
\usepackage{graphicx, color}
\usepackage{dcolumn}
\usepackage{bm}
\usepackage{hyperref}
\hypersetup{
setpagesize=false,
 bookmarksnumbered=true,%
 bookmarksopen=true,%
 colorlinks=true,%
 linkcolor=blue,
 citecolor=blue,
 urlcolor = blue,
}

\newcommand{\RM}[1]{\mathrm{#1}}
\newcommand{\etal}{\ {\it{et al.}}}

\begin{document}

\preprint{APS/123-QED}

\title{Anomalous fluctuations in homogeneous fluid phase of active Brownian particles}

\author{Yuta Kuroda}
\email{kuroda@r.phys.nagoya-u.ac.jp}
\author{Hiromichi Matsuyama}

\author{Takeshi Kawasaki}

\author{Kunimasa Miyazaki}
\email{miyazaki@r.phys.nagoya-u.ac.jp}
 
\affiliation{
 Department of Physics, Nagoya University, Nagoya 464-8602, Japan
}

\date{\today}

\begin{abstract}
Giant number fluctuations (GNF) are
an anomaly
universally observed  in active fluids 
with polar or nematic order.
In this paper, we
show that 
GNF arise in the fluid phase of active Brownian particles (ABP),
where the polar order is absent.
GNF in
ABP
extends over a large but finite length which characterizes  the growing velocity correlations.
To 
suppress unwanted phase separation and
allow ones to
 explore the disordered fluid phase at large activities,
we impart the inertia, or the mass, to the ABP.
A linearized hydrodynamic theory captures our findings, but only qualitatively.
We find numerically a
 nontrivial scaling relation for the density correlation function,
 which the linearized theory cannot explain.
The results suggest ubiquitousness of the anomalous fluctuations even
 in the disordered homogeneous fluid phase in the absence of  the
directional order. 
\end{abstract}

\maketitle

\section{introduction}
Active matter refer{s} to a broad class of many-body systems 
consisting of self-propelling constituents, such as flocks of birds,
herds of animals, bacterial colonies,
or even self-propelled colloidal particles~\cite{Bechinger2016RMP,Marchetti2013RMP,Ramaswamy2010Anuual_Review}. 
In the past few decades, we have witnessed tremendous progress in the studies
of active matter. 
Active matter systems exhibit many nontrivial phenomena that are
prohibited in equilibrium systems. Representative examples include anomalous increases of
particle number fluctuations
known as giant number fluctuations (GNF)~\cite{Ramaswamy2003EPL,Chate2006PRL,Narayan2007Science},
spatiotemporal chaotic patterns of velocities 
fields
reminiscent of
turbulence~\cite{Dombrowski2004PRL, Wensink14308}, 
and spontaneous separation of constituent particles into dense and dilute phases
called the motility-induced phase separation (MIPS)~\cite{Taillerur2008PRL,fily2012PRL}.  

The active Brownian particles (ABP) model is one of the simplest 
models of active matter~\cite{fily2012PRL}
and has been used
to study MIPS
theoretically~\cite{Stenhammar2013PRL, Bialk2013, Speck2014PRL, Speck2015,
Wittkowski2014NatCom, Cates2015Annual_Review,
RednerPRL2016,Solon2018PRE, dePirey2019PRL} and numerically~\cite{fily2012PRL, Redner2013PRL, Fily2014soft_matter,
Stenhammar2014soft_matter, Levis2017SoftMatter,
Siebert2018PRE,Digregorio2018PRL, Caporusso2020PRL}. 
MIPS resembles the liquid-vapor phase separation in equilibrium systems,
and some efforts were made to understand MIPS by mapping ABP and other
active fluids into the effective equilibrium
system~\cite{Speck2014PRL,Cates2015Annual_Review,Farage2015PRE,RednerPRL2016,Solon_2018,Speck2021pre}.   
Recently, however, it has been realized that MIPS of ABP is accompanied
by
intrinsically nonequilibrium phenomena, such as negative surface
tension~\cite{Bialke2015PRL,Solon_2018}, reversal of the Ostwald
process~\cite{Tjhung2018PRX,Shi2020PRL}, and spatial  velocity
correlation~\cite{Caprini2020PRL}.  
In particular, the spatial velocity correlation is not only observed
inside the MIPS phase but also in the high-density regimes, including the 
crystalline~\cite{Caprini2020PRR,Caprini2021Soft_Matter}, 
amorphous~\cite{Flenner2016SoftMatter,Henkes2020Nature_Communications},
and even dense fluid phases~\cite{Szamel2021EPL, Caprini2020PRR, Keta2022}.
The spatial velocity correlation is manifested as the vortex-shaped
patterns, which suggests a deep connection with the
active turbulence \cite{Keta2022}.

Since the longitudinal part of the velocity field is directly
related to the density field,
it is natural to expect that the growth of the velocity
correlation leads to an increase of the density fluctuations
similar to GNF 
in the ordered phase of polar active fluids
\cite{TonerPRL1995,Toner2005,Chate2008PRE,Marchetti2013RMP}. 
If such large density fluctuations exist in ABP fluids, it is
tempting to see the connection between them 
and GNF observed in the systems with polar long-range order. 
Several studies have reported large number fluctuations in
ABP~\cite{fily2012PRL,Fily2014soft_matter,Digregorio2018PRL}, but
 it is difficult to judge
whether the observed data are due to {\it bona fide}  
GNF or originated from heterogeneities by MIPS.

In this paper, we demonstrate that 
the homogeneous fluid state of ABP, 
despite the absence of the polar 
or nematic order, develops the large density
fluctuations and GNF, whose sizes increase 
with the growing spatial correlation of the longitudinal velocity.
The main obstacle to 
observing number fluctuations is heterogeneous density modulation caused by MIPS at high activity.
One way to avoid MIPS is 
to explore the high-density
region \cite{Caprini2020PRR,Szamel2021EPL,Keta2022}, 
but the glassy slow
dynamics or the precursor of crystallization would intervene there.
Another route is to study the intermediate-density fluid phase outside the
binodal region. However, the activity is too low to observe any meaningful signal of
the growing fluctuations.
To overcome these practical issues, 
we consider ABP with the inertia term or the mass.
It is known that MIPS is suppressed if the inertia
term is added to the original overdamped ABP \cite{Mandal2019PRL}. 
If the mass is sufficiently large, then the system remains
homogeneous without a sign of phase separation
even at high activity.
It enables one to investigate
intrinsically nonequilibrium fluctuations without being impeded by
unwanted MIPS.

We confirm numerically that the spatial velocity correlation 
develops even at intermediate densities. 
Their longitudinal and transverse modes are characterized by two distinct correlation lengths, as reported in the high-density fluid state \cite{Szamel2021EPL}.  
The transverse velocity correlation is associated with the vortex
structure, reminiscent of active turbulence \cite{Keta2022}, whereas
the longitudinal one
is accompanied by the spatial correlation of the density fluctuations.
The correlation lengths of the longitudinal velocity and
density increase with the activity.
This results in the emergence of GNF.
Contrary to the case of polar
fluids where GNF arise due to the polar order
~\cite{TonerPRL1995,Toner2005,Chate2008PRE,Marchetti2013RMP},
GNF in ABP are confined in a large but finite length scale 
corresponding to longitudinal correlation length. 
We develop
a linearized fluctuating hydrodynamic
theory from the microscopic model
and 
show that the growing lengths and GNF
can be qualitatively captured by the linearized theory.
The theory clarifies the similarities
and differences of the mechanism of GNF between our system and the 
polar active fluids in the ordered phase
\cite{TonerPRL1995, Ramaswamy2003EPL, Toner2005, Marchetti2013RMP}.
However, the linearized theory fails to explain the nontrivial scaling
relation and scaling exponents shown by the simulation.
This implies that the nonlinear coupling of fluctuations is at play.

This paper is organized as follows. In Sec.~\ref{sec2}, we describe
the model and simulation setting.
Numerical results are shown in
Sec.~\ref{sec3}. 
The analysis based on the linearized hydrodynamic theory is
sketched in Sec.~\ref{sec4}.
We devote 
Sec.~\ref{sec5} to a summary.

\begin{figure}[t]
\centering
  \includegraphics[width=9cm]{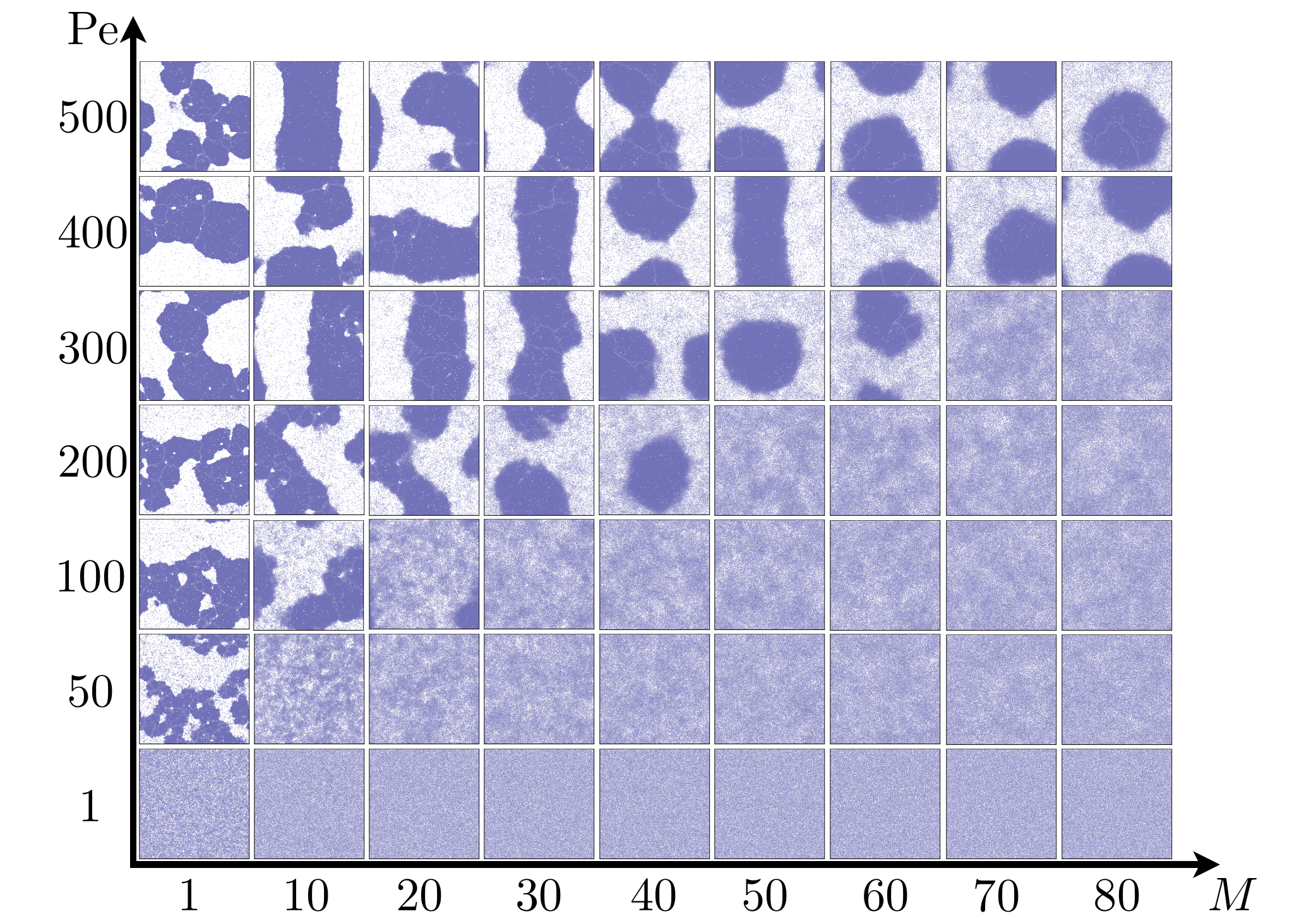}
 \caption{\label{p-d}Snapshots of particle configurations in
 $(M,\RM{Pe})$ space. $\rho=0.5$ and $N=4 \times 10^4$.
MIPS is suppressed as $M$ increases. 
In this study, we focus on the region $M=80,\ \RM{Pe} \leq 200$.
}
\end{figure}
\begin{figure*}[t]
\centering
  \includegraphics[width=18cm]{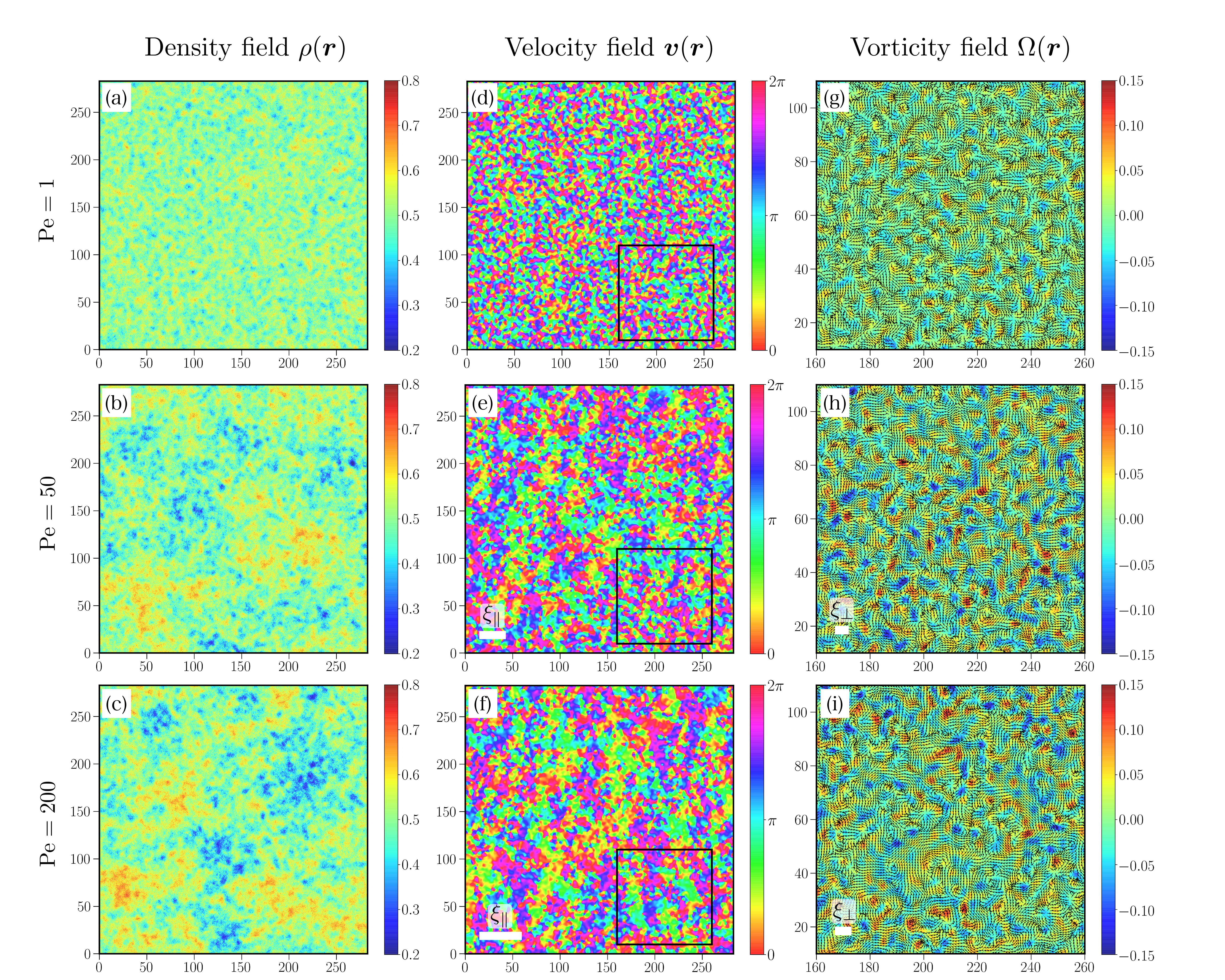}
 \caption{\label{field}
{Snapshots of} density [(a)-(c)], velocity [(d)-(f)], and vorticity fields [(g)-(i)] for $\RM{Pe}=1,50,200$ at $M=80$. Number of particles is $N=4\times10^4$.
 The horizontal and vertical axes denote the $x$ and $y$ coordinates, respectively. 
Small boxes in  panels (d)-(f) correspond to the plot range of panels (g)-(i), respectively.
The colors represent the local density in panels (a)-(c), the angle of local velocity with respect to the $x$ axis in panels (d)-(f), and the local vorticity in panels (g)-(i), respectively.
 Black arrows in panels (g)-(i) represent the direction of local velocity.
The horizontal white scale
 bars in panels (e), (f), (h), and (i)  denote the longitudinal and
 transverse correlation lengths obtained by the velocity correlation
 functions (see the text).
}
\end{figure*}
\begin{figure}[t]
\centering
\hspace{-0.7cm}
  \includegraphics[width=9.2cm]{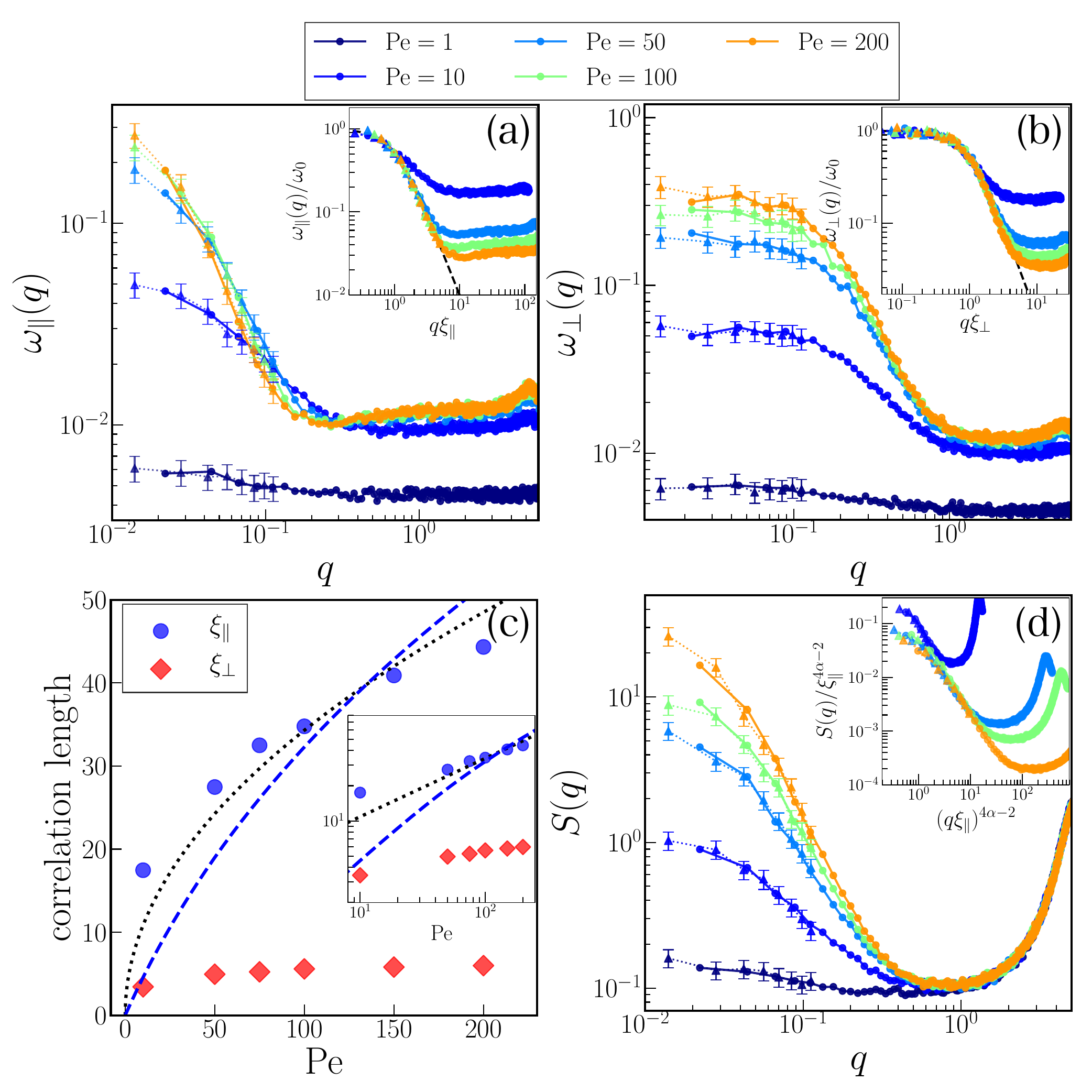}
 \caption{\label{v_corr}
The velocity correlation functions 
and static structure factor at $M=80$ in the Fourier space.
(a) The longitudinal, (b) transverse part of the velocity
 correlation functions, and (d) the static structure factor. 
Filled circles and triangles represent data for 
$N=4\times 10^4$ and $N=1\times 10^5$, respectively. 
The latter data are depicted with standard error.
{The insets} of panels (a) and (b) are the fits by the Ornstein-Zernike function (the dashed line). 
{(c)}
{The longitudinal} and transverse correlation lengths $\xi_\parallel$ and
 $\xi_\perp$ obtained by the fitting as a function of the P\'eclet number.
The dotted line is the fit by $\xi_\parallel\propto\RM{Pe}^{1/2}$, and the blue dashed line is the
fit by the linearized hydrodynamic theory (see Sec.~ \ref{sec4}).
{The inset is the log-log plot of the same data.}
{The inset of panel (d) is the rescaled curves of $S(q)$ by
 $\xi_{\parallel}$ {and $\alpha$} (see the text).} 
}
\end{figure}

\section{Model and Simulation setting}
\label{sec2}
We consider two-dimensional active Brownian particles with a finite mass, which we
 refer to as the inertial active Brownian particles (iABP).
The Langevin equation which iABP obey
is written as
\begin{align}
m\dv[2]{\bm r_j(t)}{t} &= -\zeta \dv[]{\bm r_j(t)}{t} -\nabla_j\sum_{k<l}U(r_{kl}) 
+ \zeta v_0\bm e(\phi_j),
\label{eom}
\end{align}
where $\bm r_j$ is the position of the
$j$th particle.
$m$ is the mass of a particle, $\zeta$ is the friction coefficient, {and}
$U(r_{kl})$ is the pairwise interaction potential between the particles
$k$ and $l$. $r_{kl} = |\bm r_k -\bm r_l |$ is the distance between the
two particles, and
$\nabla_j$ denotes the gradient acting on $\bm r_j$.
The last term of the right-hand side of Eq.~(\ref{eom}) is the active
noise. Its strength is characterized by the self-propelling speed $v_0$,
and the direction is {represented} by a unit vector  $\bm e(\phi_j)=(\cos\phi_j,\sin\phi_j)$.
The dynamics of orientation $\phi_j$ of the particle $j$ is described by
\begin{equation}
\dv{\phi_j(t)}{t} = \sqrt{\frac{2}{\tau_\RM{p}}} \eta_j(t) ,
\label{phi}
\end{equation}
where $\eta_j(t)$ is the Gaussian white noise that satisfies $\expval{\eta_j(t)}=0$ and $\expval{\eta_j(t)\eta_k(t')} = \delta_{j,k}\delta(t-t')$.
The symbol $\expval{\cdots}$ denotes the ensemble average. 
$\tau_\RM{p}$ is the persistence time, an essential parameter
characterizing how far the system is from equilibrium.
In the $\tau_\RM{p}\rightarrow 0$ limit, Eq.~(\ref{eom}) becomes the equilibrium Langevin equation with the effective temperature $T_\RM{eff} = v_0^2\tau_\RM{p}\zeta/2$.
Mandal\etal~\cite{Mandal2019PRL} employed the iABP with both 
the translational thermal noise and the rotational inertial term for
$\phi_j$, which we do not consider here for simplicity.

The simulation setting is as follows.
We employ the Weeks-Chandler-Andersen potential as a pairwise potential~\cite{WCA}:
\begin{equation}
U(r_{kl}) = 4\epsilon\left\{ \left( \frac{\sigma}{r_{kl} }\right)^{12} - \left( \frac{\sigma}{r_{kl} }\right)^{6} + \frac{1}{4} \right\} \theta( 2^{1/6}\sigma -r_{kl} ), 
\end{equation}
where $\theta(x)$ is the Heaviside step function,
and $\sigma$ is the diameter of a particle. 
We choose $\tau_\RM{v}= \sigma/v_0$ and
$\sigma$ as the units of time and length, respectively. 
The number density is {set relatively low at $\rho = 0.5$,
}  and the system size is $L=\sqrt{N/\rho}$. 
Control parameters in the simulation are the P\'eclet number defined by
$\RM{Pe} = {\tau_\RM{p}} /{\tau_\RM{v}}= {\tau_\RM{p}v_0} / {\sigma}$, the dimensionless mass  $M = m /(\zeta {\tau_\RM{v}})$, {and the energy ratio $\epsilon/(\zeta v_0 \sigma)$}. Here we set $\epsilon/(\zeta v_0 \sigma)=100$.
We carry out the Brownian dynamics simulation for 
iABP with the periodic boundary condition.
To integrate the equation of motion, we use the Euler-Maruyama method with a time step $\Delta t= 10^{-2} \tau_\RM{v}$.
The number of particles $N=1\times 10^4,\ 4\times 10^4$, and $\ 1\times 10^5$
are chosen to check the system size effect. 
{For the computation of the correlation functions discussed below, we
take the time average after confirming that the system is sufficiently
relaxed to the stationary state by monitoring the time evolution of the potential
energy.}

\section{Numerical results}
\label{sec3}
Figure~\ref{p-d} shows snapshots of particle configurations in $(M,\RM{Pe})$ space.
At $M=1$, the inertia effect is negligible, and the system undergoes MIPS
at $\RM{Pe} \gtrsim 50$, as reported for the overdamped ABP~\cite{fily2012PRL,Redner2013PRL,Digregorio2018PRL}.
As $M$ increases, the MIPS phase boundary line shifts to a larger $\RM{Pe}$ continuously, and at the largest $M \simeq 80$, the system remains in the homogeneous fluid phase even at $\RM{Pe}=300$ (see also Appendix~\ref{apeA} for the system size dependence).
Now that we successfully generated a homogeneous fluid with large
P\'eclet  numbers, we explore the properties of nonequilibrium
fluctuations of ABP without being intervened by unwanted inhomogeneity
induced by MIPS.

Figure~\ref{field} presents typical snapshots of the density field
$\rho(\bm r)$ [(a)-(c)], velocity field $\bm v(\bm r)$ [(d)-(f)], and
vorticity field {$\Omega(\bm r)=(\nabla \times {\bm v}(\bm r))_z$} [(g)-(i)]
for $\RM{Pe} = 1, 50$, and $200$ at $M=80$ (see Appendix~\ref{apeB} for the computation method).   
First, we focus on the velocity and vorticity fields. 
The colors in Figures~\ref{field}~(d), (e), and (f) represent the angle of vector $\bm v(\bm r)$ with respect to the $x$-axis.
{The velocity pattern is uniform for $\RM{Pe}=1$, where the system is close to equilibrium.}
As $\RM{Pe}$ increases, velocity-aligned domains appear and their sizes grow.
Concomitantly, the vorticity field $\Omega (\bm r)$ develops as
shown in Figures~\ref{field}~(g), (h), and (i). 
The sizes of the patterns, however, are appreciably smaller than those
of $\bm v (\bm r)$.  
To quantify these spatial patterns, we define the longitudinal and
transverse velocity correlation function{s} in the Fourier space
by~\cite{Szamel2021EPL}  
\begin{equation} \label{omega}
\omega_\parallel(q) = \frac{1}{N}\expval*{|J_\parallel(\bm q)|^2},\ \ \ 
\omega_\perp(q) = \frac{1}{N}\expval*{|J_\perp(\bm q)|^2}.
\end{equation}
{Here, we decomposed the Fourier transformed current $\bm J(\bm q)= \sum_{j}\dot{\bm {r}}_j e^{-i\bm q\cdot \bm r_j}$ as
 $\bm J(\bm q) = J_\parallel(\bm q)\hat{\bm q}_\parallel + J_\perp(\bm q)\hat{\bm q}_\perp$. 
$\hat{\bm q}_\parallel$ and $\hat{\bm q}_\perp$ denote the unit vector parallel and perpendicular to the wave vector $\bm q$, respectively. 
 As we can directly derive from Eq. (\ref{eom}), both $\omega_\parallel(q)$ and $\omega_\perp(q)$ take the value $\omega_0=\RM{Pe}/[2(M+\RM{Pe})]$ at $q=0$. This value is used 
for the fitting to evaluate correlation lengths discussed below.
$\omega_\parallel(q)$ is a good measure to probe the extent of the alignment
of the velocity of particles, whereas $\omega_\perp(q)$ probes the
development of the vorticity pattern. 
Figures~\ref{v_corr}~(a) and (b) show the $q$-dependence of $\omega_\parallel(q)$ and
$\omega_\perp(q)$ for various $\RM{Pe}$ at $M=80$ (see Appendix~\ref{apeE} for $\rho$ dependence). 
Both $\omega_\parallel(q)$ and $\omega_\perp(q)$ grow significantly at small wave number.
This behavior indicates the development of the {spatial} correlations
of both the longitudinal and transverse {velocities}. 
We extract correlation lengths by fitting with the Ornstein-Zernike
function $\omega_\mu(q) = {\omega_{0}}/({1+(\xi_\mu q)^2}),\ (\mu =\parallel, \perp)$,  
for the two correlation functions (see {the} insets of Figures~\ref{v_corr}~(a) and (b)) 
{\cite{Szamel2021EPL}}.
The fitting range is {$q<0.06$} for $\omega_\parallel(q)$  
and {$q<0.3$} for $\omega_\perp(q)$.
Figure~\ref{v_corr}~(c) 
shows the correlation length 
obtained by fitting for $\RM{Pe}\geq 10$.
We left out the data for $\RM{Pe}=1$ because the data are too small to extract the correlation length.
We find that the two correlation lengths are distinct;
the longitudinal length $\xi_{\parallel}$ is much longer than the
transverse counterpart $\xi_\perp${,}
and $\xi_{\parallel}$ grows
 with Pe, whereas the dependence of $\xi_\perp$ on $\RM{Pe}$ is much weaker,
which is
again consistent with the results in Ref.~\cite{Szamel2021EPL}. 
$\xi_{\parallel}$ and $\xi_{\perp}$ are comparable to 
the sizes of patterns of the velocity and the vorticity shown in
Figures~\ref{field}~(e), (f), (h), and (i).
These observations are qualitatively consistent with the numerical results
by Szamel and Flenner~\cite{Szamel2021EPL}, and the
{prediction}
of the linearized fluctuating hydrodynamic theory
by Marconi\etal~\cite{marconi2021}. 
{Note that large spatial velocity correlations can be confirmed 
in real space, as reported in
Refs. \cite{Caprini2020PRR,Caprini2021Soft_Matter}.
The
thus-obtained correlation length is close to the value of $\xi_\perp$ (see Appendix~\ref{apeC}).  
This is natural, as $\xi_\perp$ is smaller than $\xi_\parallel$.}
The vortex pattern in Figures~\ref{field}~(h) and (i) and 
{the bahavior of $\omega_{\perp}(q)$} at high $\RM{Pe}$ are reminiscent of the active
turbulence reported in various active matter
systems~\cite{Dombrowski2004PRL,
Wensink14308,Wensink_2012,Dunkel2013PRL,
Nisiguchi2015PRE,Creppy2015PRE,Guillamat:2017,Lin:2021tr, liu2020,
qi2021, alert2021}.   
We {find that} the energy spectrum $E(q)$
 obtained from the velocity correlation function
 $\omega(q) = \omega_\parallel(q) +\omega_\perp(q)$
 exhibits weak power-law behavior (see Appendix~\ref{apeD}).
However, the power-law exponent of $E(q)$ is small compared with
those reported in
{other}
studies~\cite{Dombrowski2004PRL,Wensink14308,Wensink_2012,Dunkel2013PRL,Nisiguchi2015PRE,Creppy2015PRE,Guillamat:2017,Lin:2021tr,liu2020,qi2021,alert2021}.
Seeking a link between the active turbulence and the 
{observed} {spatial}
correlation is out of the scope of the present study and {is} left for future work.

In Figure~\ref{v_corr}~(d), we show the density correlation function, or
the static structure factor, 
defined by $S(q) = \expval{\delta\rho(\bm q)\delta\rho ( - \bm q)}/N$,
where $\delta \rho(\bm q) = \rho(\bm q) -\expval{\rho(\bm q)}$ is the
fluctuation{s} of the Fourier transformed density field $\rho(\bm q) = \sum_{j}e^{-i\bm q\cdot \bm r_j}$.
$S(q)$ at small wave number is almost constant at $\RM{Pe}=1$ but 
rises significantly as $\RM{Pe}$ increases, meaning  that the density fluctuations increase at large scales.  
Note that {the} increase of $S(q)$ at small wave numbers is distinct
from that {observed} in the MIPS phase {(see Appendix~\ref{apeF})}.
In the latter case, $S(q)$ is well fitted by $q^{-(d+1)}$ ($d$ is the spatial dimension), which is called Porod's law~\cite{onuki2007, Bray2002}, 
and it is a natural consequence of the 
domains created by the phase
separation. 
On the contrary, the system in our study is spatially uniform and 
the increase of $S(q)$ observed in Figure~\ref{v_corr}~(d) is
induced by the {large} correlation of the longitudinal 
velocity field.

\begin{figure}[t]
\centering
\hspace{-0.5cm}
  \includegraphics[width=8.9cm]{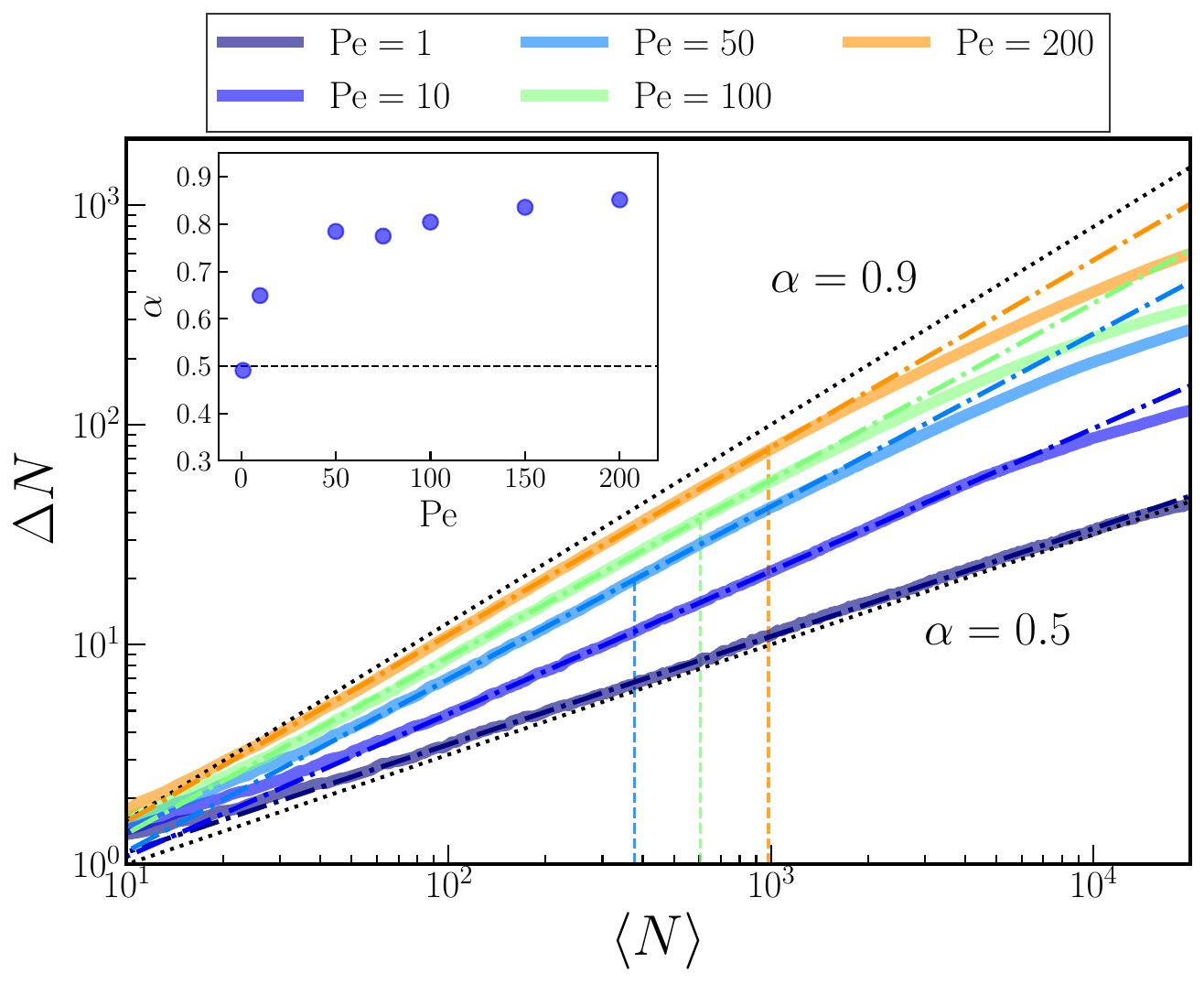}
 \caption{\label{gnf}Number fluctuation $\Delta N$ as a function of 
$\expval{N}$ at $M=80$.
{The solid} and dot-dash{ed} lines are 
simulation data {and power-law fits}, respectively.
The dotted lines indicate $\expval{N}^\alpha$ with $\alpha=0.5$ and $0.9$ shown as a guide to the eye.
{The vertical} dash{ed} 
lines represent the value of $\expval{N}$ at $\ell = \xi_\parallel$.
{The dependence of the exponent $\alpha$ on $\RM{Pe}$ is plotted in the inset.}
}
\end{figure}
Finally, we investigate the particle number fluctuations defined by
$\Delta N = \sqrt{\expval{(N-\expval{N})^2}}$.
We measure 
$\Delta N$ and the average number of particles $\expval{N}$ in the sub-box with the 
side length {$\ell$ ($<L$)}
in the whole system. 
In equilibrium systems, $\Delta N$ should be proportional to $\expval{N}^{1/2}$.
In active matter {with polar or nematic order}, however, GNF
characterized by $\Delta N \propto \expval{N}^{\alpha}$ with a larger exponent $\alpha
>0.5$ {are} observed~\cite{Narayan2007Science,Chate2008PRE,Zhang2010PNAS,Chate2006PRL,Ginelli2012PRL, Peruani2012PRL,Nao2014PRL,Nishiguchi2017PRE,Kawaguchi:2017,Mahault2019PRL, Iwasawa2021PRR}.   
The iABP model is ideal for examining GNF because MIPS is absent even at large P\'eclet numbers.
In Figure~\ref{gnf},  we plot $\Delta N$ as a function of $\expval{N}$ for several $\RM{Pe}$ for a fixed $M(=80)$.
$\Delta N$ behaves as $\expval{N}^\alpha$ with exponent
{$\alpha>0.5$} for large {$\RM{Pe}$}.
We chose the fitting range as $\expval {N}\in[100,1000]$ to extract the exponent $\alpha$. 
The dot-dashed lines in Figure~\ref{gnf} are the power-law fit of the simulation data.
Interestingly, the {side length}
{$\ell$}
at which $\Delta N$ deviates from the power-law {is comparable to} $\xi_\parallel$, as indicated by vertical dashed lines in Figure~\ref{gnf}. 
The dependence of the exponent $\alpha$ on $\RM{Pe}$ is plotted in the inset of Figure~\ref{gnf}.
Starting from the smallest value of $\alpha \simeq 0.5$ at $\RM{Pe}=1$,
$\alpha$ increases with $\RM{Pe}$, up to {$\alpha \simeq 0.85$} at the
largest $\RM{Pe}$.

The number fluctuation $\Delta N$ is related to the static
structure factor $S(q)$ by $S(q \rightarrow 0) = \Delta N^2/\expval{N}$ at large ${\ell}$.  
{Therefore, $\Delta N \propto \expval{N}^{\alpha}$
leads to {$S(q) \propto q^{-\beta}$ in the reciprocal space.}}
{T}he two exponents are related 
by $\beta = 4\alpha -2$~\cite{Ramaswamy2003EPL,Ginelli:2016vx}.
On the other hand, 
it is natural to expect
that the density fluctuation{s} {are} characterized by the
correlation length {of the longitudinal 
 {velocity correlation}  function}, $\xi_{\parallel}$.
Thus, we assume the scaling form
\begin{equation}
\label{5}
S(q) = \xi_{\parallel}^{\beta}f(q\xi_{\parallel} ),
\end{equation}
where the scaling function satisfies $f(x)
\sim {\it const.}$ for $x<1$ and $f(x) \sim x^{-\beta}$ for $x>1$.
The inset of Figure~\ref{v_corr}~(d) is the rescaled plot of $S(q)$ for $\RM{Pe}\geq 10$.
The data collapse to a single curve for $\RM{Pe}\geq 50$, but the data for $\RM{Pe}=10$ deviates from the curve.
This 
{supports} the validity of the scaling ansatz at least for $\RM{Pe}\geq 50$
and confirms the relation between GNF and $S(q)$.

\section{Qualitative description of Giant Number Fluctuations}
\label{sec4}

{
To explain the connection between
the velocity correlation and GNF observed above, here we develop a
linearized fluctuating hydrodynamic theory for the homogeneous fluid
state of iABP. 
For active fluids with polar or nematic order, the linearized
hydrodynamics explain GNF as a result of coupling between the density field and order parameter, which is prohibited in equilibrium systems
\cite{Ramaswamy2003EPL, Toner2005, Marchetti2013RMP}. 
GNF in the 
fluid state of 
iABP, where the order or Goldstone
modes are absent, arise by a similar but different
mechanism. 
In this section, we sketch their derivation. 
Following Dean's method~\cite{Dean_1996, Nakamura_2009} 
and assuming that the interaction term (pressure gradient)
linearly depends only on the density, we can derive 
the linearized equation for the density, current, and polarization fields
from Eqs.~(\ref{eom}) and (\ref{phi}) (see Appendix~\ref{apeG} for derivation):
\begin{equation}
\begin{aligned}
\partial_t \delta\rho(\bm r,t) &= -\nabla \cdot \delta\bm J(\bm r,t),
\\
m \partial_t{\delta \bm J(\bm r,t)}
&= - \frac{1}{\rho\chi}\nabla\delta\rho(\bm r,t) - \zeta\delta \bm J(\bm r,t)  + \zeta v_0 \delta \bm p(\bm r ,t), 
\\
\partial_t{\delta \bm p(\bm r,t)} &= -\frac{1}{\tau_\RM{p}}\delta \bm p(\bm r,t) 
+ \sqrt{\frac{\rho}{\tau_\RM{p}}} \bm \Upsilon(\bm r,t),  
\label{6} 
\end{aligned}
\end{equation}
where  $\bm p(\bm r,t) =\sum_{j=1}^{N}\bm e(\phi_j(t))\delta(\bm r-\bm
r_j(t))$ denotes the polarization,
and $\bm \Upsilon(\bm r,t)$
is the Gaussian white noise with zero mean and the correlation
$\expval{ \Upsilon _\alpha(\bm r,t)\Upsilon _\beta(\bm
r',t')}=\delta_{\alpha,\beta}\delta(\bm r-\bm r')\delta(t-t')$
with $\alpha, \beta = x,y$.
The coefficient $\chi$ is the ``compressibility''. 
From Eq.~(\ref{6}), it is straightforward to calculate the equal time correlation functions in Fourier space. 
The longitudinal velocity correlation function $\omega_\parallel(q)$ and static structure factor $S(q)$ are calculated as
\begin{equation}
\label{hh-39}
\omega_\parallel(q) = \frac{\omega_0}{1+(\xi_\parallel q)^2}
\end{equation}
and 
\begin{equation}
\label{hh-42}
S(q) 
=\frac{S_0}{1+(\xi_\parallel q)^2},
\end{equation}
respectively (see Appendix~\ref{apeG}). Here,
the values at $q=0$ are given by $\omega_0= {v_0^2\tau_\RM{p}}/({2(\tau_\RM{m} + \tau_\RM{p})})$
and 
$S_0 ={\rho\zeta\chi v_0^2}/{2D}$.
$\xi_\parallel=\sqrt{{\tau_\RM{p}}/[\rho\zeta\chi({1+\tau_\RM{m}/\tau_\RM{p}})]}$
is the correlation length.
The theory predicts that $\omega_{\parallel}(q)$ and $S(q)$ 
are characterized by the same correlation length $\xi_\parallel$, which supports
numerical results shown in Figures~\ref{v_corr}(a) and (d).
Furthermore, Eq.~(\ref{hh-42}) means that the density correlation function behaves as $S(q)\sim q^{-2}$ 
on length scales smaller than $\xi_\parallel$.
From the argument above Eq.~(\ref{5}), this yields 
GNF; $\Delta N\sim \expval{N}^{1}$ with the exponent $\alpha=1$. 
The argument given above elucidates how
GNF in our system arise due to
the growth of the spatial
longitudinal velocity correlation caused by persistence motion, and
they are confined in the region of size $\xi_\parallel$.
This also explains the numerical results in Figure~\ref{gnf} qualitatively.
We note that 
the linearized hydrodynamic theory can explain the growth of the
correlations of
the density and
longitudinal velocity, but it cannot predict
the growth of the transverse velocity or the vortex, as pointed out  in Ref.~\cite{Szamel2021EPL}. 
}

Finally, we remark that  the
prediction of the linear hydrodynamic theory is only qualitative.
Recall that the static structure factor $S(q)$ 
satisfies the scaling relation Eq.(\ref{5}) with
the exponent $\beta=4\alpha -2$ and {$\beta$} 
varies with $\RM{Pe}$ (cf. the inset of Figure~\ref{gnf}).
In contrast, the linearized theory predicts the Ornstein-Zernike form with the fixed $\beta(=2)$.
Also, $S(q)$ obtained numerically
is larger than predicted by the
linearized theory 
at small wave numbers
(see Appendix~\ref{apeG}).  
These observations suggest that nonlinear coupling of the
fluctuations  between different hydrodynamic modes is at play.
Furthermore, as shown by the blue dashed line in Figure \ref{v_corr}(c), the fit by theoretical 
prediction of $\xi_\parallel$ (below Eq.~(\ref{h-42})) deviates from the numerical data.
Note that, in the small $M$ limit, our theoretical prediction for $\xi_\parallel$ is reduced to $\xi_\parallel \propto \RM{Pe}^{1/2}$ obtained theoretically for overdamped ABP \cite{Szamel2021EPL}. Somehow the fit by $\xi_\parallel \propto \RM{Pe}^{1/2}$ (dotted line in Figure \ref{v_corr}(c)) works better than our theoretical prediction. More quantitative assessments of these results are left for feature work.

\section{summary}
\label{sec5}
{
In this paper, we
studied the growing density or number fluctuations in the disordered
homogeneous phase of ABP for a wide range of P\'eclet numbers.
It was possible by introducing the inertia to the original overdamped ABP model, 
which
suppresses MIPS and generates the disordered homogeneous fluid. 
This system is ideal
for studying the inherent
nonequilibrium fluctuations unimpeded by MIPS.
We first confirmed that the spatial velocity correlation has two
distinct correlation lengths, $\xi_\parallel$ and $\xi_\perp$, corresponding to the longitudinal and
transverse modes, even at a relatively low density.
$\xi_\perp$ corresponds to the size of vortex patterns, which is reminiscent of active turbulence in the simple spherical active matter \cite{Keta2022}.
$\xi_\parallel$ 
is longer than  $\xi_\perp$ and grows with the P\'eclet number. 
The growing longitudinal velocity correlation
is related 
to the spatial correlation of the density fluctuation or the structure
factor $S(q)$. 
We found that $S(q)$ grows at small
wave numbers
with P\'eclet numbers and has the same characteristic length as the longitudinal velocity correlation function.
The large density fluctuations in the wave vector space 
is nothing but the large number fluctuations, or GNF, in the real space. 
We measured the number fluctuation $\Delta N$ in a sub-box of the size $\ell$ and
showed that it grows 
as $\Delta N \sim \langle N\rangle^{\alpha}$
with the exponent $\alpha > 0.5$.  The exponent $\alpha$ increases
monotonically with P\'eclet number. 
The largest sub-box size below which we observe 
GNF
agrees with $\xi_\parallel$. 
These facts yield a scaling relation for $S(q)$ characterized by $\xi_\parallel$ and $\alpha$.
Our results provide a coherent picture of the origin of  GNF
observed in ABP. 
We address that the origin of GNF here is similar 
but strictly different from GNF observed in ordered active fluids~\cite{Chate2008PRE,Mahault2019PRL}.
In the ordered active fluids such as the Vicsek model,
the active nematic~\cite{ Chate2006PRL, Nao2014PRL}, or
the self-propelled rods~\cite{ Ginelli2012PRL}, 
GNF are understood as the
``infection'' of Goldstone modes of the ordered phase to the density
field~\cite{TonerPRL1995, Ramaswamy2003EPL, Toner2005,
Marchetti2013RMP}. 
In our model, however, the system is globally disordered, and there
is no Goldstone mode.
Instead, the large spatial velocity correlation
yields GNF. 
We also showed that our results can be captured by a linearized hydrodynamic theory qualitatively but not quantitatively.
A quantitatively valid theoretical treatment would require an analysis
that fully 
considers nonlinear
couplings of fluctuations.
Our results suggest that 
the anomalously large density fluctuations should be universally
and ubiquitously present in various active matter systems,  
even without explicit global orders and phase separation. 
}

\begin{acknowledgments}
We thank Daiki Nishiguchi, Kazumasa A. Takeuchi, Kyosuke Adachi,
Ludovic Berthier, and Yann-Edwin Keta for fruitful discussions.
This work was supported by KAKENHI 18H01188, 19H01812, 19K03767, 20H05157, 20H00128, JST SPRING (Grant Number JPMJSP2125), and JST FOREST Program (Grant Number JPMJFR212T).
The authors (YK and HM) thank the ``Interdisciplinary Frontier Next-Generation Researcher Program of the Tokai Higher Education and Research System."
\end{acknowledgments}

\appendix
\section{System size dependence of phase behavior }
\label{apeA}
\begin{figure}[h]
\centering
  \includegraphics[width=9cm]{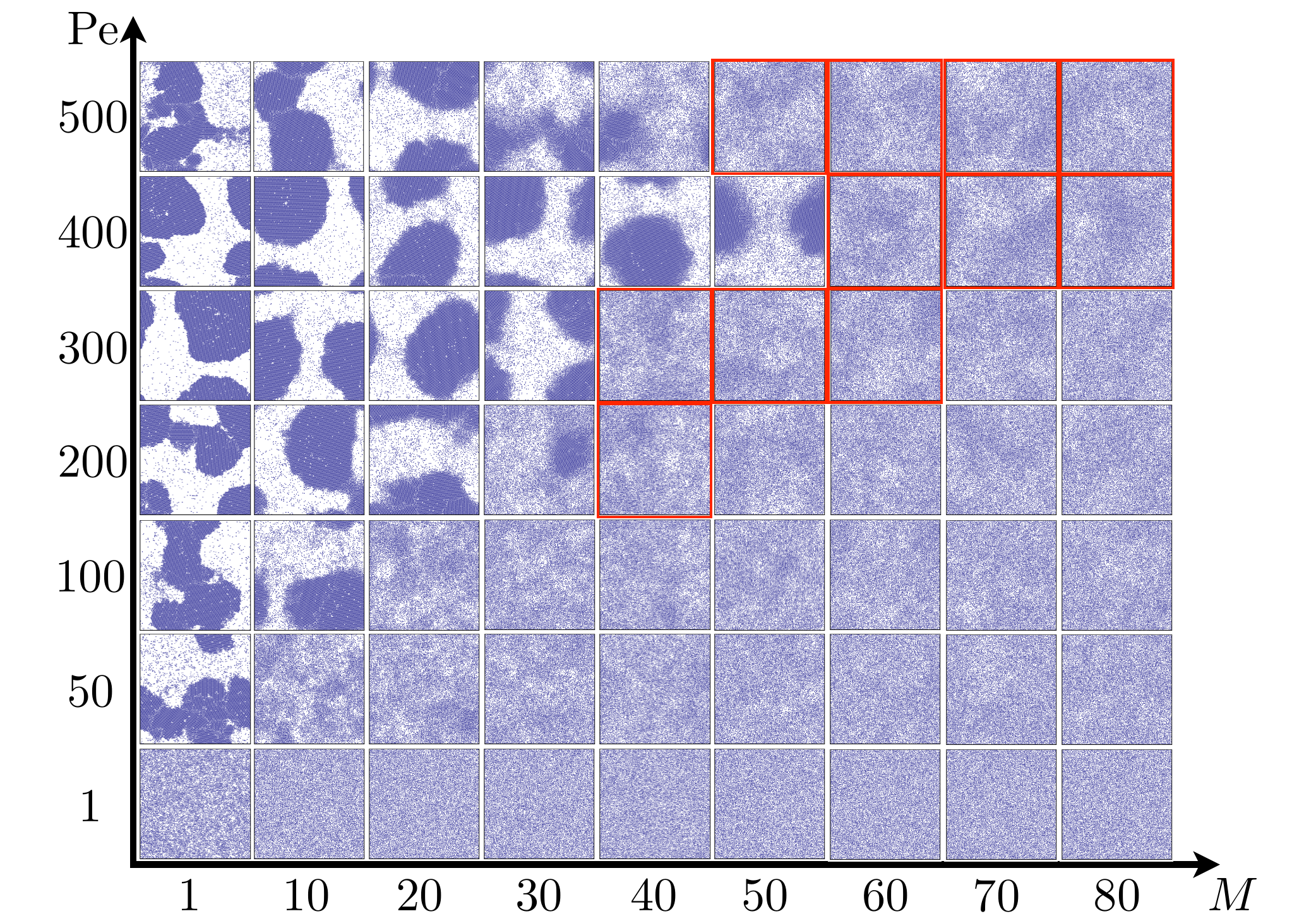}
 \caption{\label{10000}The ``phase diagram'' of iABP for the system size $N=1\times 10^4$. 
Each panel is the snapshot of particle configuration for corresponding
 parameters $(M, \rm{Pe})$.
 For the parameters indicated by the red-colored panels, the system undergoes MIPS for larger system size $N=4\times 10^4$.
 }
\end{figure}
\begin{figure*}[t]
\centering
  \includegraphics[width=14cm]{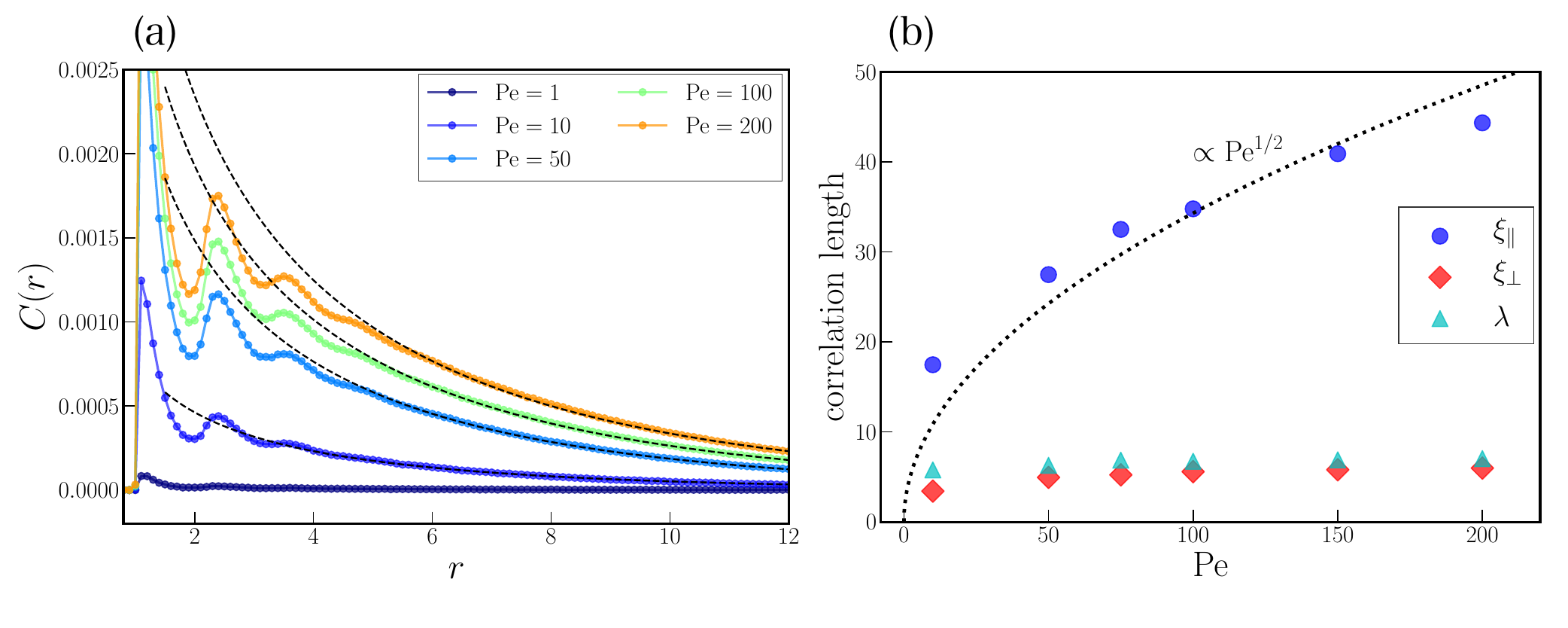}
 \caption{\label{cr}
(a) The velocity correlation function in the real space. 
The mass is fixed at $M=80$. 
The dashed lines are 
fits 
by $C(r) = Ar^{-1/2}e^{-r/\lambda}$.
 (b) {The correlation} lengths of {the} velocity correlation function. 
{The triangle symbols are the correlation length obtained from $C(r)$.
The circles and diamonds are $\xi_\parallel$ and $\xi_\perp$ shown in
 Figure~\ref{v_corr}(c) in the main text. 
}
The dotted line is a fit by $\xi_\parallel\propto \RM{Pe}^{1/2}$.
}
\end{figure*}
In Figure~\ref{p-d} of the main text, we have shown the ``phase diagram'' in
$(M, \rm{Pe})$ space and the phase boundary between the MIPS and homogeneous
phase. 
We have chosen a relatively large simulation size of $N=4\times 10^4$
because the phase boundary is sensitive to the system size. 
In Figure~\ref{10000}, we show the phase diagram obtained from the
smaller system $N=1\times 10^4$.
The red-colored panels are configurations for the parameters
in which the system undergoes MIPS at
a larger system size of $N=4\times 10^4$, as shown in Figure~\ref{p-d} of the main text.
Furthermore, {Figure~\ref{10000}
shows}  that MIPS disappears at very large
{Pe} ($\geq 500$) {in the small system}. 
This re-entrant transition
is reminiscent of the results shown by Mandal\etal~\cite{Mandal2019PRL}, in which the inertia of both the position and the rotation (of the
active noises) as well as the thermal noise are taken into account. 
We address that {the} re-entrance 
{observed in our current model}
is the artifact due to the small system
size. 

\section{Calculation of coarse-grained density, velocity, and vorticity
 fields} 
 \label{apeB}
We have shown the coarse-grained density, velocity, and vorticity fields in Figure~\ref{field} of the main text.
These quantities are calculated as  follows.
The local density $\rho(\bm r)$ is obtained by 
averaging the number of particles in a circle with {a} radius {of}
$3\sigma$ placed on every node of a square-lattice with the lattice constant $\sigma$.
The velocity field $\bm v(\bm r)$ was obtained by taking the Gaussian-weighted average 
in a circle with {a} radius {of} $3\sigma$.
The value of the variance of the Gaussian function is chosen in such a way
that the Gaussian function is $0.1$ at $r=3\sigma$.
The vorticity field $\Omega(\bm r) = \partial_x v_y - \partial_y v_x$ 
is calculated as  $\Omega(\bm r) \simeq \sum_{\RM{cell}}\bm v (\bm r)\cdot \delta \bm r / \delta S_{\RM{cell}}$, 
{a line-integral along the circumference of a square cell with a side length $0.25\sigma$. 
$\delta S_{\RM{cell}}$ is the area of the cell.}

\section{Velocity correlation function in real space}
\label{apeC}
In Figure \ref{cr}(a), we show the velocity correlation function in real
space \cite{Caprini2020PRR, Caprini2021Soft_Matter}, that is defined by
\begin{equation}
C(r) = \frac{1}{N}\expval{\sum_{j\neq k} \bm v_j\cdot \bm v_k \delta(\bm r- \bm r_j + \bm r_k)}.
\end{equation}
This is 
the Fourier transformation of
$\omega(q) = \omega_\parallel(q)+ \omega_\perp(q)$ {introduced in the main text}. 
One observes that the spatial velocity
correlation {grows} {as Pe increases} in the real space.
{We fit the data by a function $C(r) = Ar^{-1/2}e^{-r/\lambda}$, which is 
the Fourier transformation of the Ornstein-Zernike function at large $r$~\cite{zwillinger2014}.}
{The dashed} lines in Figure \ref{cr}(b) are {fits} by this function. 
The fitting range is chosen as $r>5$ for all Pe{'s}.
The triangle symbol{s} in Figure \ref{cr}~(b) represent
the correlation length $\lambda$ that is found by the fitting of $C(r)$. 
{We find} that $\lambda \simeq \xi_{\perp}$
and $C(r)$ is dominated by the transverse part{.} {This is natural} because $\xi_\perp \ll
\xi_\parallel$ as seen in Figure \ref{cr}~(b). 

\section{Energy spectrum and velocity distribution}
\label{apeD}
\begin{figure*}[t]
\centering
  \includegraphics[width=14cm]{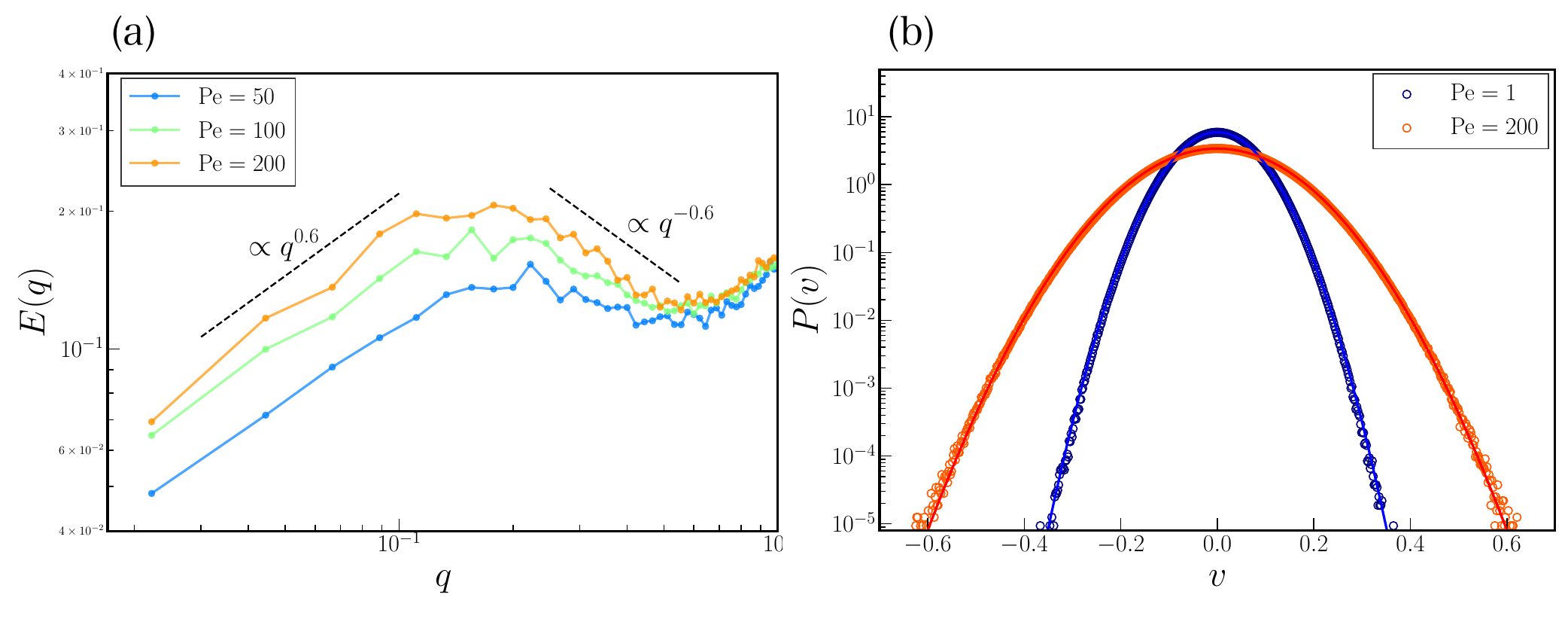}
 \caption{\label{v_dist}(a) {The energy} spectrum $E(q)$ for
 {Pe$=50,~100$}, and $200$ as a function of $q$. 
The mass is fixed at $M=80$. 
The black broken line{s} of $q^{-0.6}$ {and $q^{0.6}$} {are} guide{s} for the eyes. 
 (b) {The velocity} distribution function $P(v)$ for $M=80$. 
 The {empty circles} with blue and
 orange edge color{s} denote the 
 numerical results {for} $\RM{Pe}=1$ and $200$, respectively.  
 {The solid} lines are fits by the Gaussian
 distribution. 
 }
\end{figure*}
\begin{figure*}[t]
\centering
  \includegraphics[width=18cm]{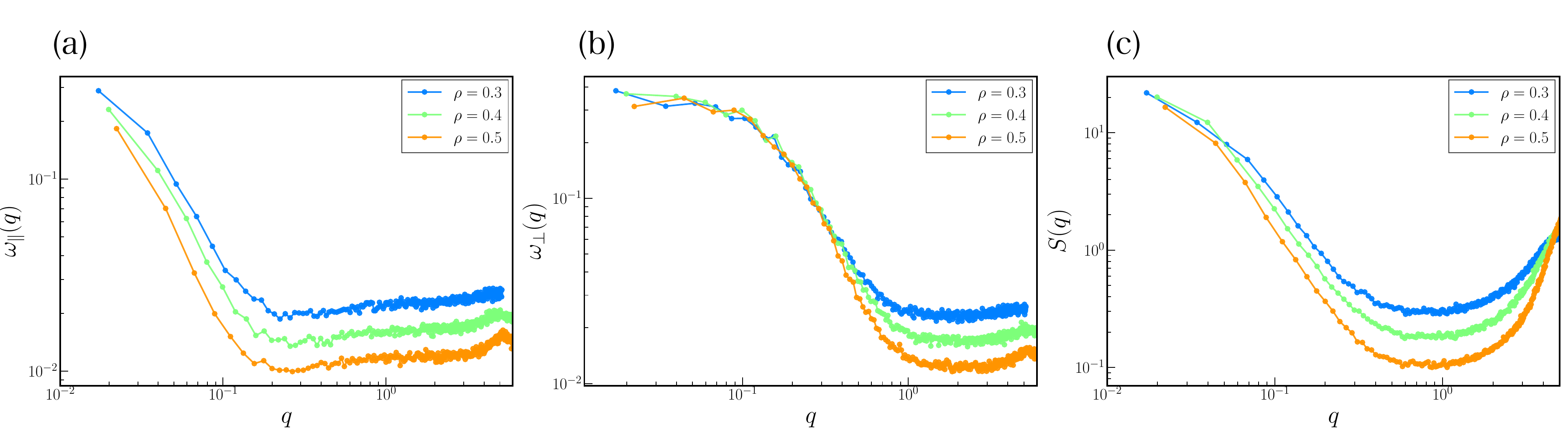}
 \caption{\label{density} {Density dependence of spatial correlation
 functions at $\RM{Pe}=200$. The mass is fixed at $M=80$. 
(a) The longitudinal velocity
 correlation function, (b) transverse velocity correlation function, and
 (c) static structure factor for $\rho=0.3,0.4,$ and
 $0.5$. All data are from the simulation with $N=4\times 10^4$.
 }
}
\end{figure*}
In the main text, we observed the development of the vortex structures 
whose spatial patterns are reminiscent of the turbulence. 
In the standard inertial turbulence of fluids at high Reynolds numbers, 
the fingerprint of the turbulence is the 
universal scale-free behavior of the energy spectrum, known as the
Kolmogorov law~\cite{frisch_1995}.
A similar power law is also found in the energy spectrum in various
active matter
systems~\cite{Dombrowski2004PRL,Wensink14308,Wensink_2012,
Dunkel2013PRL, Nisiguchi2015PRE,Creppy2015PRE,Guillamat:2017,
Lin:2021tr, liu2020, qi2021, alert2021}. 
Such behaviors are aptly called active turbulence. 
However, the exponent of the power law depends on systems. Little is known about the universality of  active turbulence.
Here we show the energy spectrum of the {i}ABP model studied in
the main text. 
In two dimensions, the energy spectrum is related to the velocity
correlation function $\omega(q)=\omega_\parallel(q)+\omega_\perp(q)$ by 
\begin{equation}
E(q) = 2\pi q\omega(q). 
\end{equation}
In Figure~\ref{v_dist}~(a), the energy spectra $E(q)$ for several \rm{Pe}'s at 
$M=80$ are shown. 
One observes a faint sign of the power law with the amplitudes
increasing with \rm{Pe} at intermediate
wave numbers at $q \gtrsim 0.1$. 
A crude estimate of 
the exponent $\gamma$ of the power law $E(q) \sim q^{-\gamma}$ is
{approximately equal to}
 0.6, which is
much smaller than {values} reported in the
past~\cite{Dombrowski2004PRL,Wensink14308,Wensink_2012,Dunkel2013PRL,
Nisiguchi2015PRE,Creppy2015PRE,Guillamat:2017,Lin:2021tr,liu2020,qi2021,
alert2021}.    

Recently, the non-Gaussianity of the velocity distribution has been
reported in ABP and the active Ornstein-Uhlenbeck {particles} (AOUP) 
at high densities and high \rm{Pe}~\cite{Caprini2020jcp,Keta2022}.  
We evaluated the velocity distribution to check if such deviation is
also observed for low densities.
In Figure~\ref{v_dist}(b), we show the velocity distribution for
$\RM{Pe} = 1$ and 200 at $M=80$. 
The solid lines are the corresponding Gaussian distribution defined by 
\begin{equation}
P(v)=\sqrt{\frac{M}{2\pi T_\mathrm{kin}}}\exp\left( -\frac {Mv^2}{2T_\mathrm{kin}}\right),
\end{equation}
where $T_\mathrm{kin} = M\expval{v_x^2+v_y^2}/2$ is the kinetic temperature.
For both \rm{Pe}'s, the observed distribution functions are 
well fitted by the Gaussian,
{as in other systems at turbulent states~\cite{Wensink14308, Dunkel2013PRL, qi2021}.}


\section{Density dependence of spatial correlations}
\label{apeE}
In the main text,  we showed spatial correlation functions only at $\rho=0.5$.
However, these large spatial correlations exist even at more low densities. 
Figure \ref{density} represents spatial correlation functions for $\rho=0.3,0.4,$ and $0.5$. 
We confirm existence of large correlations for all quantities, longitudinal velocity, transverse velocity, and density correlation functions.
Hence, we
conclude that the results in the main text are insensitive to the
densities.

\begin{figure}[t]
\centering
  \includegraphics[width=8.7cm]{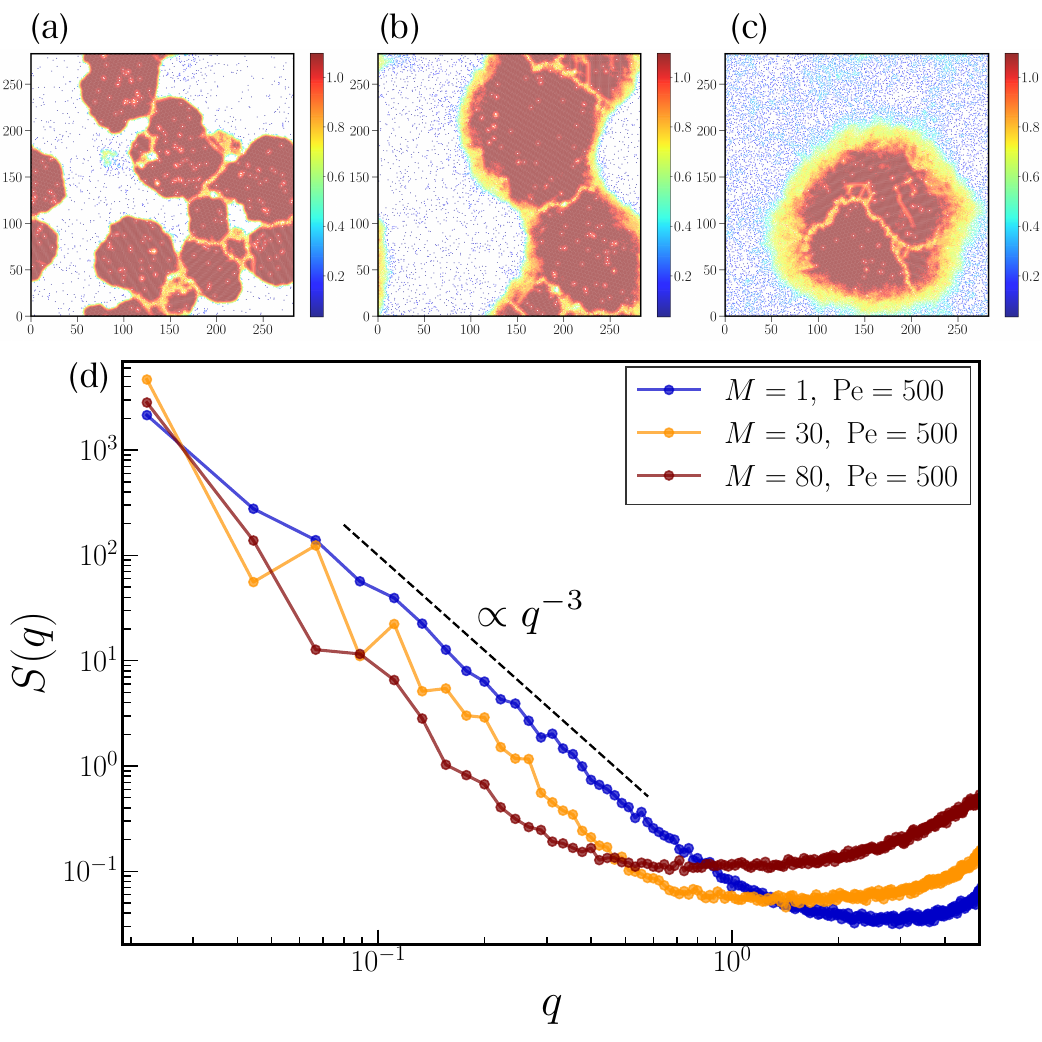}
 \caption{\label{MIPS}
{Snapshots in MIPS phases} for {$\RM{Pe}=500$}, (a)$M=1$, (b)$M=30$, and (c)$M=80$.
The {color bar indicates the magnitude of density.}
 {The system size} is $N=4\times 10^4$. 
{(d)} The static structure factor $S(q)$ for each parameter.}
\end{figure}

\section{Density correlation  in MIPS phase}
\label{apeF}
It is known that the system undergoing the phase separation 
with smooth surfaces develops the peak in the static structure factor
characterized by a power-law,
$S(q)\propto q^{-(d+1)}$ in the low-wave-number regime. 
This is called Porod's law~\cite{onuki2007, Bray2002}. 
Porod's law is also observed in MIPS {phase} of active
matter~\cite{Stenhammar2014soft_matter,Caporusso2020PRL}.  
We show that Porod's law is also observed for iABP when the system undergoes MIPS. 
Figure~\ref{MIPS}(a), (b), and (c) are snapshots of the system
undergoing MIPS for several $M$'s.
The color{s} represent the local density
calculated by averaging the number of particles in a
circle with {a} radius {of} $3\sigma$.
For the smallest inertia, $M=1$, the phase boundary is sharp, and their surface is smooth. 
When $M=30$ and $80$, on the other hand, the phase {boundaries} become
diffusive, and the surfaces are blurred. 
This behavior might be related to the difference 
{in} {the}
(effective) temperatures between the dense and gas phase in the
presence of inertia~\cite{Mandal2019PRL}.
In Figure~\ref{MIPS}~(d), we show the static structure factor $S(q)$ for
$M=1,~30${,} and 80.
For $M=1$, Porod's law, {\it i.e}, $S(q)\propto q^{-3}$, is clearly
observed at low $q$'s~\cite{Stenhammar2014soft_matter, Caporusso2020PRL}.
For the higher inertia, $M=30$ and $80$, $S(q)$ deviates from Porod's
law, although the heights of $S(q)$ at low $q$'s are unaltered. 
This behavior should be the consequence of the change in the sharpness
of the phase boundaries.

In any case, we address that the development of the peak of $S(q)$ at
low $q$'s reported in the main text is distinct from trivial Porod's
law of MIPS.
\begin{widetext}
\section{Fluctuating hydrodynamic description}
\label{apeG}
\label{hydro}

{
In Sec.~\ref{sec4} of the main text, we employed an effective hydrodynamic description to elucidate the qualitative mechanism of the large density fluctuations or GNF.
Here, we derive an effective hydrodynamic equation for iABP by following Dean's method \cite{Dean_1996, Nakamura_2009}, and calculate the longitudinal velocity correlation function and static structure factor.
}

\subsection{Derivation of the fluctuating hydrodynamic equations} 

\label{deri}
Our stating point is the equation of motion for 
the inertial active Brownian particles (iABP) in two dimension:
\begin{align}
\dv{\bm r_j(t)}{t} &= \bm v_j(t) ,\\
m\dv{\bm v_j(t)}{t} &= -\zeta \bm v_j(t) - \sum_{k=1}^{N}\nabla_j
 U(r_{jk}) + \zeta v_0\bm e(\phi_j(t)) \label{h-2},\\
\dv{\phi_j(t)}{t} &= \sqrt{\frac{2}{\tau_\RM{p}}}\eta_j(t) \label{h-3}.
\end{align}
Here, $\eta_j(t)$ is a white noise that satisfies $\expval{\eta_j(t)}=0$
 and $\expval{\eta_j(t)\eta_k(t')}=\delta_{j,k}\delta(t-t')$.
$U(r)$ is the pairwise potential. 
We assume that $\nabla U(0)= \bm 0$ for simplicity. 
$\bm e(\phi)=(\cos\phi,\sin\phi)$ is the unit vector pointing to the
 direction of the active random force. 
Hydrodynamic fields of this system are the number density
\begin{equation}
\label{h-4}
\rho(\bm r,t) = \sum_{j=1}^{N}\delta(\bm r- \bm r_j(t)),
\end{equation}
density current
\begin{equation}
\label{h-5}
\bm J(\bm r,t) = \sum_{j=1}^{N}\bm v_j(t) \delta(\bm r- \bm r_j(t)),
\end{equation}
and polarization
\begin{equation}
\label{h-6}
\bm p(\bm r,t) = \sum_{j=1}^{N}\bm e(\phi_j(t)) \delta(\bm r- \bm r_j(t)).
\end{equation}
By differentiating these hydrodynamic fields with respect to time,
we obtain the following set of equations. 
For the density, it is the continuum equation;
\begin{equation}
\partial_t \rho(\bm r,t) = -\nabla \cdot \bm J(\bm r,t) .
\end{equation}
For the current and the polarization fields, 
\begin{align}
m \partial_t{\bm J(\bm r,t)}
&= -\nabla\cdot \mathsf M^\RM{vv}(\bm r,t)  -\zeta \bm J(\bm r,t) - \rho(\bm r,t)\sum_{k=1}^{N}\nabla U(\bm r-\bm r_k) + \zeta v_0 \bm p(\bm r ,t),  \label{h-8}\\
\partial_t {\bm p(\bm r,t)} &= - \nabla\cdot \qty( \mathsf M^\RM{ev}(\bm r,t))^\RM{T} - \sum_{j=1}^{N}\dv{\bm e(\phi_j(t))}{t}\delta(\bm r-\bm r_j(t)),   \label{h-9}
\end{align}
where tensors $\mathsf M^\RM{vv}(\bm r,t)$ and  $\mathsf M^\RM{ev}(\bm
r,t)$ are defined by
\begin{align}
 \mathsf M^\RM{vv}(\bm r,t) &:= m\sum_{j=1}^{N}  \bm v_j(t)\bm v_j(t) \delta(\bm r-\bm r_j(t)) \label{h-10},\\
  \mathsf M^\RM{ev}(\bm r,t) &:= \sum_{j=1}^{N}\bm e(\phi_j(t))\bm v_j(t)\delta(\bm r-\bm r_j(t)).\label{h-11}
\end{align}
These tensors can be rewritten in terms of hydrodynamic fields, 
following the procedure discussed in Ref.\cite{Nakamura_2009}, 
as
\begin{align}
\mathsf M^\RM{vv}(\bm r,t) &= \frac{m \bm J(\bm r,t)\bm J(\bm r,t)}{\rho(\bm r,t)}, \\
\mathsf M^\RM{ev}(\bm r,t) &= \frac{\bm p(\bm r,t)\bm J(\bm r,t)}{\rho(\bm r,t)}.
\end{align}
The potential part in the right-hand side of Eq.~(\ref{h-8}) can be expressed as
\begin{equation}
\sum_{k=1}^{N} \nabla U(\abs{\bm r-\bm r_k}) 
=
\nabla\fdv{\mathcal F[\rho(\cdot,t)]}{\rho(\bm r,t)},
\end{equation}
where {the} functional $\mathcal F[\rho]$ is defined by
\begin{equation}
\mathcal F[\rho(\cdot,t)] := \frac{1}{2}\int_{V}\dd[2]\bm r\int_{V}\dd[2]\bm r'
\ \rho(\bm r,t)\rho(\bm r',t)U(\abs{ \bm r-\bm r' }).
\end{equation}
Substituting these expressions, Eq.~(\ref{h-8}) becomes 
\begin{equation}
m \partial_t{\bm J(\bm r,t)}
= -\nabla\cdot \qty(\frac{m \bm J(\bm r,t)\bm J(\bm r,t)}{\rho(\bm r,t)} )  -\zeta \bm J(\bm r,t) - \rho(\bm r,t) \nabla\fdv{\mathcal F[\rho(\cdot,t)]}{\rho(\bm r,t)} + \zeta v_0 \bm p(\bm r ,t) .
\end{equation}
Next, we derive the equation for polarization. The time derivative of the unit vector $\bm e(\phi_j(t))$ in right-hand side of Eq.~(\ref{h-9}) is given by 
\begin{align}
\dv{\bm e(\phi_j(t))}{t} 
&= \sqrt{\frac{2}{\tau_\RM{p}}}\mqty(-\sin\phi_j(t)  \\ \cos\phi_j(t) ) \circ\eta_j(t) \notag \\
&= -\frac{1}{\tau_\RM{p}}\bm e_j(t) 
+  \sqrt{\frac{2}{\tau_\RM{p}}}\mqty(-\sin\phi_j(t)  \\ \cos\phi_j(t) )   \bullet \eta_j(t),
\label{h-17}
\end{align}
where the symbols $\circ$ and $\bullet$ denote
the Stratonovich and It\^o product, respectively.  
We have adopted the It\^o representation for the multiplicative
noise to ensure that the average of the noise is zero~\cite{gardiner}.
Using Eq.~(\ref{h-17}), 
Eq.~(\ref{h-9}) is rewritten as
\begin{equation}
\partial_t{\bm p(\bm r,t)} = -\frac{1}{\tau_\RM{p}}\bm p(\bm r,t) 
- \nabla\cdot \qty(  \frac{\bm J(\bm r,t)\bm p(\bm r,t)}{\rho(\bm r,t)}) + \bm \Lambda(\bm r,t),
\end{equation}
where the noise term $\bm \Lambda(\bm r,t)$ is defined as
\begin{equation}
\label{h-20}
\bm \Lambda(\bm r,t):= \sqrt{\frac{2}{\tau_\RM{p}}}\sum_{j=1}^{N}\mqty(-\sin\phi_j(t)  \\ \cos\phi_j(t) )   \bullet \eta_j(t) \delta(\bm r-\bm r_j(t)).
\end{equation}
We rewrite this noise as
\begin{equation}
\label{h-21}
\bm \Lambda(\bm r,t)=\sqrt{\frac{\rho(\bm r,t)}{\tau_\RM{p}}} \bm \Upsilon(\bm r,t),
\end{equation}
which satisfies $\expval{\Upsilon_\alpha(\bm r,t)}=0$ and 
\begin{equation}
\label{h-21b}
\expval{\Upsilon_\alpha(\bm r,t)\Upsilon_\beta(\bm r' , t') } = \delta_{\alpha,\beta}\delta(\bm r -\bm r')\delta(t-t').
\end{equation}
We can prove Eq.~(\ref{h-21b}), by calculating the each component of 
noise correlations and compare the results from Eq.~(\ref{h-20}).
For example, the $(x, x)$ component is calculated as
\begin{align}
\label{h-22}
\expval{\Lambda_x(\bm r,t)\Lambda_x(\bm r',t')} 
&=  \frac{2}{\tau_\RM{p}}\sum_{j=1}^{N}\expval{\sin^2\phi_j(t)}
\delta(\bm r-\bm r_j(t))\delta(\bm r -\bm r')\delta(t-t').
\end{align}
Using Eq.~(\ref{h-3}), the expectation value of $\sin^2\phi_j(t)$ can be obtained as
\begin{equation}
\expval{\sin^2\phi_j(t)} = \frac{1}{2} - \frac{1}{2}\cos(2\phi_j(0))e^{-4t/\tau_\RM{p}}.
\end{equation}
The summation $\sum_{j=1}^{N}\cos(2\phi_j(0))$ becomes $0$ in the limits of $N\rightarrow \infty$ because the initial value of angles $\phi_j(0)$ is completely random.
Hence, in the limit of $N\rightarrow \infty$, Eq.~(\ref{h-22}) becomes
\begin{equation}
\label{h-24}
\expval{\Lambda_x(\bm r,t)\Lambda_x(\bm r',t')}  =  \frac{\rho(\bm r,t)}{\tau_\RM{p}} \delta(\bm r -\bm r')\delta(t-t') .
\end{equation}
The $(y,y)$ component of Eq.~(\ref{h-20})
is also given by Eq.~(\ref{h-24}) in the limit of $N\rightarrow \infty$. 
The correlation function between $x$ and $y$ component of Eq.~(\ref{h-20}) becomes $0$ by using the relation
\begin{equation}
\expval{\sin\phi_j(t)\cos\phi_j(t)} = \frac{1}{2}\sin(2\phi_j(0))e^{-4t/\tau_\RM{p}}.
\end{equation}
It is noteworthy that the noise correlation for the polarization field
is identical to those of another, or simpler, active matter model
known as the active Ornstein-Uhlenbeck (AOUP) model~\cite{Fodor2016PRL}, in the continuum limit.

Below, we summarize the derived fluctuating hydrodynamic equation for iABP;
\begin{align}
\partial_t \rho(\bm r,t) &= -\nabla \cdot \bm J(\bm r,t),\label{h-26} \\
m \partial_t{\bm J(\bm r,t)}
&= -\nabla\cdot\mathsf P(\bm r,t) - \zeta \bm J(\bm r,t)  -\nabla\cdot \qty(\frac{m \bm J(\bm r,t)\bm J(\bm r,t)}{\rho(\bm r,t)} )   + \zeta v_0 \bm p(\bm r ,t), \\
\partial_t{\bm p(\bm r,t)} &= -\frac{1}{\tau_\RM{p}}\bm p(\bm r,t) 
- \nabla\cdot \qty(  \frac{\bm J(\bm r,t)\bm p(\bm r,t)}{\rho(\bm r,t)}) + \sqrt{\frac{\rho(\bm r,t)}{\tau_\RM{p}}} \bm \Upsilon(\bm r,t), \label{h-28}
\end{align}
with the pressure tensor $\mathsf P(\bm r,t)$ defined by 
\begin{equation}
\label{h-29}
\nabla\cdot\mathsf P(\bm r,t) := \rho(\bm r,t) \nabla\fdv{\mathcal F[\rho(\cdot,t)]}{\rho(\bm r,t)}.
\end{equation}

Now let us consider the linearization of the fluctuating hydrodynamics
of iABP so that we can derive the 
correlation functions. 
We assume that the pressure tensor Eq.~(\ref{h-29}) depends 
{only on}
the density field.  
Up to the linear order in the density fluctuation 
$\delta \rho(\bm r,t) = \rho(\bm r,t) - \rho$ in the 
hydrodynamic limit~\cite{Szamel2021EPL}, we have
\begin{equation} 
\label{h-29b}
\nabla\cdot\mathsf P(\bm r,t) \simeq \frac{1}{\rho\chi}\nabla \delta
 \rho(\bm r,t).
\end{equation}
Here $\chi$ is a ``compressibility'' defined by ${\chi}^{-1} :=
\rho\eval{ \partial{P}/\partial{\rho} }_{\rho(\bm r)=\rho} $ 
where $P$ is a diagonal component of $\mathsf P(\bm r,t)$. 
Linearizing Eqs.~(\ref{h-26})-(\ref{h-28}) and using 
Eq.~(\ref{h-29b}), we arrive at 
\begin{align}
\partial_t \delta\rho(\bm r,t) &= -\nabla \cdot \delta\bm J(\bm r,t),\label{h-31} \\
m \partial_t{\delta \bm J(\bm r,t)}
&= - \frac{1}{\rho\chi}\nabla\delta\rho(\bm r,t) - \zeta\delta \bm J(\bm r,t)  + \zeta v_0 \delta \bm p(\bm r ,t), \label{h-32}\\
\partial_t{\delta \bm p(\bm r,t)} &= -\frac{1}{\tau_\RM{p}}\delta \bm p(\bm r,t) 
+ \sqrt{\frac{\rho}{\tau_\RM{p}}} \bm \Upsilon(\bm r,t).  \label{h-33}
\end{align}
Note that the equation for the polarization fluctuation
Eq.~(\ref{h-33})
is a simple Ornstein-Uhlenbeck process and,  
thus, we can
{regard $\delta \bm p(\bm r,t)$ as a colored noise of
Eq.~(\ref{h-32}). In  other words, Eq.~(\ref{h-32}) is written as}
\begin{equation}
\label{h-34}
m \partial_t{\delta \bm J(\bm r,t)}
= - \frac{1}{\rho\chi}\nabla\delta\rho(\bm r,t) - \zeta\delta \bm J(\bm r,t)  +  \bm \Xi^\RM{act}(\bm r,t),
\end{equation}
{with an active noise} $\bm \Xi^\RM{act}(\bm r,t):= \zeta v_0\delta\bm p(\bm r,t)$ 
{whose correlation is written as}
\begin{equation}
\expval{\Xi^\RM{act}_\alpha(\bm r,t) \Xi^\RM{act}_\beta(\bm r',t')} 
= \frac{v_0^2\zeta^2\rho}{2}e^{-|t-t'|/\tau_\RM{p}}\delta_{\alpha,\beta}\delta(\bm r-\bm r').
\end{equation}
In the limit of $\tau_\RM{p}\rightarrow 0$, 
the active noise becomes white noise
{and the fluctuation dissipation relation is recovered.}

{Recently}, Marconi\etal\cite{marconi2021} has derived
{similar}
fluctuating hydrodynamic equations for the underdamped ABP and AOUP
model {starting from the BBGKY hierarchy}.

\subsection{Velocity and density correlation functions}
\label{corr}
From Eqs.~(\ref{h-31}) and (\ref{h-34}), we can easily calculate the
longitudinal velocity and density correlation function. By 
Fourier transforming in time and space,
Eq.~(\ref{h-31}) and (\ref{h-34}) 
are written as 
\begin{align}
-i\omega \delta\check\rho(\bm q,\omega) &= -i q \delta\check{ J}_\parallel(\bm q,\omega), 
\label{h-36} \\
-i\omega \delta\check{J}_\parallel(\bm q,\omega) &= - i\gamma b q\delta\check{\rho}(\bm q,\omega)
-\gamma \delta\check{J}_\parallel(\bm q,\omega) + \frac{1}{m}\check{\Xi}^\RM{act}_x(\bm q,\omega),
\label{h-37}
\end{align}
where $\gamma = \zeta/m$ and $b=1/(\rho\zeta\chi)$. 
The variable{s} with check symbol $\check{X}(\bm q,\omega)$ {represent} 
the Fourier transformed {quantities with} respect to $\bm r$ and $t$.
By eliminating the density field from Eq.~(\ref{h-37}) and using the Wiener-Khinchin theorem, we obtain the dynamical longitudinal velocity correlation function in Fourier space:
\begin{equation}
\label{h-38}
\omega_{\parallel}(q,\omega) =
\frac{1}{N}\int_{-\infty}^{\infty}\dd t\ 
\expval{ \delta\tilde{J_\parallel}(\bm q,t)  \delta\tilde{J_\parallel}^* (\bm q,0)}e^{i\omega t}
= \frac{v_0^2\gamma^2 D \omega^2}{[(\omega^2-\gamma bq^2)^2 +\gamma^2\omega^2 ]({\omega^2 + D^2})},
\end{equation}
where $D=1/\tau_\RM{p}$ and  
the variables with tildes, $\tilde{X}(\bm q,t)$, are 
the Fourier transformed variables with respect to $\bm r$. 
By integrating Eq.~(\ref{h-38}) over $\omega$,  we obtain 
the equal time correlation function, 
\begin{equation}
\label{h-39}
\omega_\parallel(q) = \frac{1}{2\pi}\int_{-\infty}^{\infty}\dd\omega\ \omega_{\parallel}(q,\omega) 
= \frac{\omega_0}{1+(\xi_\parallel q)^2},
\end{equation}
with
\begin{equation}
\label{h-40}
\omega_0:= \frac{v_0^2\gamma}{2(D+\gamma)} =\frac {v_0^2\tau_\RM{p}}{2(\tau_\RM{m} + \tau_\RM{p})},\ \ \ \ \ \ \ \xi_\parallel ^2:= \frac{b\gamma}{D(D+\gamma)} = \frac{b\tau_\RM{p}}{1+\tau_\RM{m}/\tau_\RM{p}}.
\end{equation}
Here, $\tau_\RM{m} = 1/\gamma$ is the inertial relaxation time. 
Next, we calculate the density correlation function. Using Eq.~(\ref{h-36}), the dynamical structure factor is written as
\begin{align}
S(q,\omega) &= \frac{q^2}{\omega^2}\omega_{\parallel}(q,\omega) 
=  \frac{v_0^2\gamma^2 D q^2}{[(\omega^2-\gamma bq^2)^2 +\gamma^2\omega^2](\omega^2 + D^2)}.
\end{align}
By integrating over $\omega$, we obtain the static structure factor
given by
\begin{equation}
\label{h-42}
S(q) = \frac{1}{2\pi}\int_{-\infty}^{\infty}\dd\omega\ S(q,\omega)
=\frac{S_0}{1+(\xi_\parallel q)^2}
\end{equation}
with
\begin{equation}
S_0 
:= \frac{v_0^2}{2b D} =\frac{1}{2}v_0^2\tau_\RM{p}\zeta\rho\chi
= \rho T_\RM{eff}\chi,
\end{equation}
where we defined the effective temperature 
by $T_\RM{eff}:= v_0^2\tau_\RM{p}\zeta/2$.
{Both $\omega_{\parallel}(q)$ and $S(q)$ are of} the Ornstein-Zernike type
are characterized by a single correlation length $\xi_\parallel$.

\begin{figure*}[t]
\centering
  \includegraphics[width=14cm]{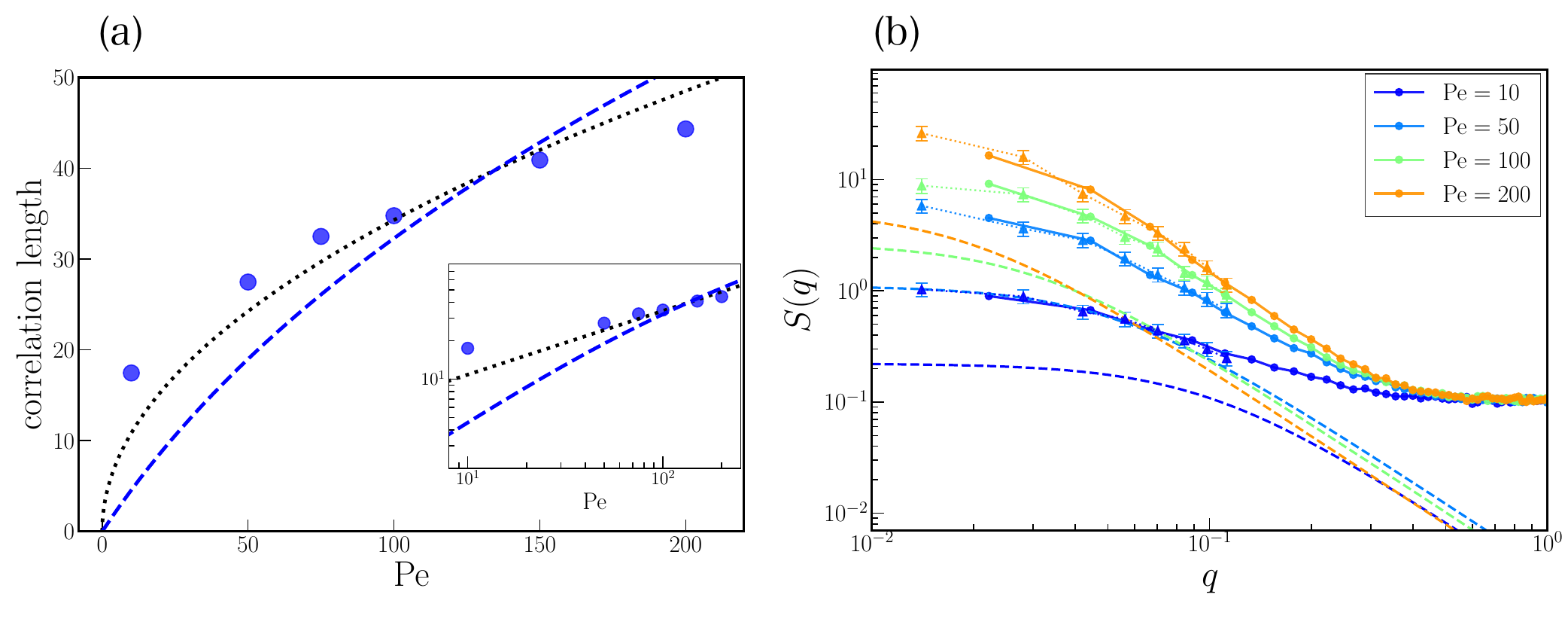}
 \caption{\label{th_BD}
(a) Numerical data and the fit by the theory [second equation of
 Eq.~(\ref{h-40})] of the longitudinal correlation length
 $\xi_{\parallel}$ (identical to Figure~\ref{v_corr}(c) in the main text).
 The inset is the log-log plot of the same data. 
The blue dots represent the numerical values. The blue dashed line is the fitting curve of the second equation of Eq.~(\ref{h-40}). 
The fitting parameter are found to be $b\tau_\RM{v}/\sigma^2\simeq$
 $18.8$. The black dotted line represents the prediction from
 the overdamped ABP, $\xi_\parallel\propto \RM{Pe}^{1/2}$. 
Panel (b)
is the numerical results of the static structure factor. 
Dashed lines represent the theoretical prediction of Eq.~(\ref{h-42}) drawn by using
 the fitting parameter
 obtained from panel (a). 
}
\end{figure*}

\subsection{Comparison of the linearized theory with numerical results}

Here we quantitatively compare simulation results to theoretical prediction.
The filled circles in Figure~\ref{th_BD}(a) are the same data for
$\xi_\parallel$ presented in Figure~\ref{v_corr}(c) in the main text.  
Recall that $\xi_\parallel$'s are obtained by fitting $\omega_{\parallel}(q)$ with the
Ornstein-Zernike function. 
We fit the data by our theoretical prediction, the second equation of
 Eq.~(\ref{h-40}),  (dashed lines) using $b\tau_\RM{v}/\sigma^2=18.8$ as a
fitting parameter.
The dotted line is $\xi_{\parallel} \propto \sqrt{\rm{Pe}}$ which was predicted by the overdamped ABP in Refs.\cite{Szamel2021EPL,Henkes2020Nature_Communications} and works better than our theoretical prediction. 
However, since the differences between the two predictions are not large, it is early to decide which scaling works better.
Using the fitting parameter obtained from Figure~\ref{th_BD} (a), 
we compare the simulated $S(q)$ with theoretical prediction, Eq.~(\ref{h-42}).
Substantial discrepancies between simulation data and theoretical
prediction can not be remedied by a slight change of the fitting
parameter $b\tau_\RM{v}/\sigma^2$ and, therefore,
implies that the nonlinear coupling of the fluctuations, which are completely absent in our theoretical analysis,
is not negligible at large \rm{Pe}'s.

\end{widetext}
%


\begin{thebibliography}{72}%
\makeatletter
\providecommand \@ifxundefined [1]{%
 \@ifx{#1\undefined}
}%
\providecommand \@ifnum [1]{%
 \ifnum #1\expandafter \@firstoftwo
 \else \expandafter \@secondoftwo
 \fi
}%
\providecommand \@ifx [1]{%
 \ifx #1\expandafter \@firstoftwo
 \else \expandafter \@secondoftwo
 \fi
}%
\providecommand \natexlab [1]{#1}%
\providecommand \enquote  [1]{``#1''}%
\providecommand \bibnamefont  [1]{#1}%
\providecommand \bibfnamefont [1]{#1}%
\providecommand \citenamefont [1]{#1}%
\providecommand \href@noop [0]{\@secondoftwo}%
\providecommand \href [0]{\begingroup \@sanitize@url \@href}%
\providecommand \@href[1]{\@@startlink{#1}\@@href}%
\providecommand \@@href[1]{\endgroup#1\@@endlink}%
\providecommand \@sanitize@url [0]{\catcode `\\12\catcode `\$12\catcode
  `\&12\catcode `\#12\catcode `\^12\catcode `\_12\catcode `\%12\relax}%
\providecommand \@@startlink[1]{}%
\providecommand \@@endlink[0]{}%
\providecommand \url  [0]{\begingroup\@sanitize@url \@url }%
\providecommand \@url [1]{\endgroup\@href {#1}{\urlprefix }}%
\providecommand \urlprefix  [0]{URL }%
\providecommand \Eprint [0]{\href }%
\providecommand \doibase [0]{https://doi.org/}%
\providecommand \selectlanguage [0]{\@gobble}%
\providecommand \bibinfo  [0]{\@secondoftwo}%
\providecommand \bibfield  [0]{\@secondoftwo}%
\providecommand \translation [1]{[#1]}%
\providecommand \BibitemOpen [0]{}%
\providecommand \bibitemStop [0]{}%
\providecommand \bibitemNoStop [0]{.\EOS\space}%
\providecommand \EOS [0]{\spacefactor3000\relax}%
\providecommand \BibitemShut  [1]{\csname bibitem#1\endcsname}%
\let\auto@bib@innerbib\@empty
\bibitem [{\citenamefont {Bechinger}\ \emph {et~al.}(2016)\citenamefont
  {Bechinger}, \citenamefont {Di~Leonardo}, \citenamefont {L\"owen},
  \citenamefont {Reichhardt}, \citenamefont {Volpe},\ and\ \citenamefont
  {Volpe}}]{Bechinger2016RMP}%
  \BibitemOpen
  \bibfield  {author} {\bibinfo {author} {\bibfnamefont {C.}~\bibnamefont
  {Bechinger}}, \bibinfo {author} {\bibfnamefont {R.}~\bibnamefont
  {Di~Leonardo}}, \bibinfo {author} {\bibfnamefont {H.}~\bibnamefont
  {L\"owen}}, \bibinfo {author} {\bibfnamefont {C.}~\bibnamefont {Reichhardt}},
  \bibinfo {author} {\bibfnamefont {G.}~\bibnamefont {Volpe}},\ and\ \bibinfo
  {author} {\bibfnamefont {G.}~\bibnamefont {Volpe}},\ }\bibfield  {title}
  {\bibinfo {title} {Active particles in complex and crowded environments},\
  }\href {https://doi.org/10.1103/RevModPhys.88.045006} {\bibfield  {journal}
  {\bibinfo  {journal} {Rev. Mod. Phys.}\ }\textbf {\bibinfo {volume} {88}},\
  \bibinfo {pages} {045006} (\bibinfo {year} {2016})}\BibitemShut {NoStop}%
\bibitem [{\citenamefont {Marchetti}\ \emph {et~al.}(2013)\citenamefont
  {Marchetti}, \citenamefont {Joanny}, \citenamefont {Ramaswamy}, \citenamefont
  {Liverpool}, \citenamefont {Prost}, \citenamefont {Rao},\ and\ \citenamefont
  {Simha}}]{Marchetti2013RMP}%
  \BibitemOpen
  \bibfield  {author} {\bibinfo {author} {\bibfnamefont {M.~C.}\ \bibnamefont
  {Marchetti}}, \bibinfo {author} {\bibfnamefont {J.~F.}\ \bibnamefont
  {Joanny}}, \bibinfo {author} {\bibfnamefont {S.}~\bibnamefont {Ramaswamy}},
  \bibinfo {author} {\bibfnamefont {T.~B.}\ \bibnamefont {Liverpool}}, \bibinfo
  {author} {\bibfnamefont {J.}~\bibnamefont {Prost}}, \bibinfo {author}
  {\bibfnamefont {M.}~\bibnamefont {Rao}},\ and\ \bibinfo {author}
  {\bibfnamefont {R.~A.}\ \bibnamefont {Simha}},\ }\bibfield  {title} {\bibinfo
  {title} {Hydrodynamics of soft active matter},\ }\href
  {https://doi.org/10.1103/RevModPhys.85.1143} {\bibfield  {journal} {\bibinfo
  {journal} {Rev. Mod. Phys.}\ }\textbf {\bibinfo {volume} {85}},\ \bibinfo
  {pages} {1143} (\bibinfo {year} {2013})}\BibitemShut {NoStop}%
\bibitem [{\citenamefont {Ramaswamy}(2010)}]{Ramaswamy2010Anuual_Review}%
  \BibitemOpen
  \bibfield  {author} {\bibinfo {author} {\bibfnamefont {S.}~\bibnamefont
  {Ramaswamy}},\ }\bibfield  {title} {\bibinfo {title} {The mechanics and
  statistics of active matter},\ }\href
  {https://doi.org/10.1146/annurev-conmatphys-070909-104101} {\bibfield
  {journal} {\bibinfo  {journal} {Annual Review of Condensed Matter Physics}\
  }\textbf {\bibinfo {volume} {1}},\ \bibinfo {pages} {323} (\bibinfo {year}
  {2010})}\BibitemShut {NoStop}%
\bibitem [{\citenamefont {Ramaswamy}\ \emph {et~al.}(2003)\citenamefont
  {Ramaswamy}, \citenamefont {Simha},\ and\ \citenamefont
  {Toner}}]{Ramaswamy2003EPL}%
  \BibitemOpen
  \bibfield  {author} {\bibinfo {author} {\bibfnamefont {S.}~\bibnamefont
  {Ramaswamy}}, \bibinfo {author} {\bibfnamefont {R.~A.}\ \bibnamefont
  {Simha}},\ and\ \bibinfo {author} {\bibfnamefont {J.}~\bibnamefont {Toner}},\
  }\bibfield  {title} {\bibinfo {title} {Active nematics on a substrate: Giant
  number fluctuations and long-time tails},\ }\href
  {https://doi.org/10.1209/epl/i2003-00346-7} {\bibfield  {journal} {\bibinfo
  {journal} {Europhysics Letters ({EPL})}\ }\textbf {\bibinfo {volume} {62}},\
  \bibinfo {pages} {196} (\bibinfo {year} {2003})}\BibitemShut {NoStop}%
\bibitem [{\citenamefont {Chat\'e}\ \emph {et~al.}(2006)\citenamefont
  {Chat\'e}, \citenamefont {Ginelli},\ and\ \citenamefont
  {Montagne}}]{Chate2006PRL}%
  \BibitemOpen
  \bibfield  {author} {\bibinfo {author} {\bibfnamefont {H.}~\bibnamefont
  {Chat\'e}}, \bibinfo {author} {\bibfnamefont {F.}~\bibnamefont {Ginelli}},\
  and\ \bibinfo {author} {\bibfnamefont {R.}~\bibnamefont {Montagne}},\
  }\bibfield  {title} {\bibinfo {title} {Simple model for active nematics:
  Quasi-long-range order and giant fluctuations},\ }\href
  {https://doi.org/10.1103/PhysRevLett.96.180602} {\bibfield  {journal}
  {\bibinfo  {journal} {Phys. Rev. Lett.}\ }\textbf {\bibinfo {volume} {96}},\
  \bibinfo {pages} {180602} (\bibinfo {year} {2006})}\BibitemShut {NoStop}%
\bibitem [{\citenamefont {{Narayan}}\ \emph {et~al.}(2007)\citenamefont
  {{Narayan}}, \citenamefont {{Ramaswamy}},\ and\ \citenamefont
  {{Menon}}}]{Narayan2007Science}%
  \BibitemOpen
  \bibfield  {author} {\bibinfo {author} {\bibfnamefont {V.}~\bibnamefont
  {{Narayan}}}, \bibinfo {author} {\bibfnamefont {S.}~\bibnamefont
  {{Ramaswamy}}},\ and\ \bibinfo {author} {\bibfnamefont {N.}~\bibnamefont
  {{Menon}}},\ }\bibfield  {title} {\bibinfo {title} {{Long-Lived Giant Number
  Fluctuations in a Swarming Granular Nematic}},\ }\href
  {https://doi.org/10.1126/science.1140414} {\bibfield  {journal} {\bibinfo
  {journal} {Science}\ }\textbf {\bibinfo {volume} {317}},\ \bibinfo {pages}
  {105} (\bibinfo {year} {2007})}\BibitemShut {NoStop}%
\bibitem [{\citenamefont {Dombrowski}\ \emph {et~al.}(2004)\citenamefont
  {Dombrowski}, \citenamefont {Cisneros}, \citenamefont {Chatkaew},
  \citenamefont {Goldstein},\ and\ \citenamefont
  {Kessler}}]{Dombrowski2004PRL}%
  \BibitemOpen
  \bibfield  {author} {\bibinfo {author} {\bibfnamefont {C.}~\bibnamefont
  {Dombrowski}}, \bibinfo {author} {\bibfnamefont {L.}~\bibnamefont
  {Cisneros}}, \bibinfo {author} {\bibfnamefont {S.}~\bibnamefont {Chatkaew}},
  \bibinfo {author} {\bibfnamefont {R.~E.}\ \bibnamefont {Goldstein}},\ and\
  \bibinfo {author} {\bibfnamefont {J.~O.}\ \bibnamefont {Kessler}},\
  }\bibfield  {title} {\bibinfo {title} {Self-concentration and large-scale
  coherence in bacterial dynamics},\ }\href
  {https://doi.org/10.1103/PhysRevLett.93.098103} {\bibfield  {journal}
  {\bibinfo  {journal} {Phys. Rev. Lett.}\ }\textbf {\bibinfo {volume} {93}},\
  \bibinfo {pages} {098103} (\bibinfo {year} {2004})}\BibitemShut {NoStop}%
\bibitem [{\citenamefont {Wensink}\ \emph {et~al.}(2012)\citenamefont
  {Wensink}, \citenamefont {Dunkel}, \citenamefont {Heidenreich}, \citenamefont
  {Drescher}, \citenamefont {Goldstein}, \citenamefont {L{\"o}wen},\ and\
  \citenamefont {Yeomans}}]{Wensink14308}%
  \BibitemOpen
  \bibfield  {author} {\bibinfo {author} {\bibfnamefont {H.~H.}\ \bibnamefont
  {Wensink}}, \bibinfo {author} {\bibfnamefont {J.}~\bibnamefont {Dunkel}},
  \bibinfo {author} {\bibfnamefont {S.}~\bibnamefont {Heidenreich}}, \bibinfo
  {author} {\bibfnamefont {K.}~\bibnamefont {Drescher}}, \bibinfo {author}
  {\bibfnamefont {R.~E.}\ \bibnamefont {Goldstein}}, \bibinfo {author}
  {\bibfnamefont {H.}~\bibnamefont {L{\"o}wen}},\ and\ \bibinfo {author}
  {\bibfnamefont {J.~M.}\ \bibnamefont {Yeomans}},\ }\bibfield  {title}
  {\bibinfo {title} {Meso-scale turbulence in living fluids},\ }\href
  {https://doi.org/10.1073/pnas.1202032109} {\bibfield  {journal} {\bibinfo
  {journal} {Proceedings of the National Academy of Sciences}\ }\textbf
  {\bibinfo {volume} {109}},\ \bibinfo {pages} {14308} (\bibinfo {year}
  {2012})}\BibitemShut {NoStop}%
\bibitem [{\citenamefont {Tailleur}\ and\ \citenamefont
  {Cates}(2008)}]{Taillerur2008PRL}%
  \BibitemOpen
  \bibfield  {author} {\bibinfo {author} {\bibfnamefont {J.}~\bibnamefont
  {Tailleur}}\ and\ \bibinfo {author} {\bibfnamefont {M.~E.}\ \bibnamefont
  {Cates}},\ }\bibfield  {title} {\bibinfo {title} {Statistical mechanics of
  interacting run-and-tumble bacteria},\ }\href
  {https://doi.org/10.1103/PhysRevLett.100.218103} {\bibfield  {journal}
  {\bibinfo  {journal} {Phys. Rev. Lett.}\ }\textbf {\bibinfo {volume} {100}},\
  \bibinfo {pages} {218103} (\bibinfo {year} {2008})}\BibitemShut {NoStop}%
\bibitem [{\citenamefont {Fily}\ and\ \citenamefont
  {Marchetti}(2012)}]{fily2012PRL}%
  \BibitemOpen
  \bibfield  {author} {\bibinfo {author} {\bibfnamefont {Y.}~\bibnamefont
  {Fily}}\ and\ \bibinfo {author} {\bibfnamefont {M.~C.}\ \bibnamefont
  {Marchetti}},\ }\bibfield  {title} {\bibinfo {title} {Athermal phase
  separation of self-propelled particles with no alignment},\ }\href
  {https://doi.org/10.1103/PhysRevLett.108.235702} {\bibfield  {journal}
  {\bibinfo  {journal} {Phys. Rev. Lett.}\ }\textbf {\bibinfo {volume} {108}},\
  \bibinfo {pages} {235702} (\bibinfo {year} {2012})}\BibitemShut {NoStop}%
\bibitem [{\citenamefont {Stenhammar}\ \emph {et~al.}(2013)\citenamefont
  {Stenhammar}, \citenamefont {Tiribocchi}, \citenamefont {Allen},
  \citenamefont {Marenduzzo},\ and\ \citenamefont {Cates}}]{Stenhammar2013PRL}%
  \BibitemOpen
  \bibfield  {author} {\bibinfo {author} {\bibfnamefont {J.}~\bibnamefont
  {Stenhammar}}, \bibinfo {author} {\bibfnamefont {A.}~\bibnamefont
  {Tiribocchi}}, \bibinfo {author} {\bibfnamefont {R.~J.}\ \bibnamefont
  {Allen}}, \bibinfo {author} {\bibfnamefont {D.}~\bibnamefont {Marenduzzo}},\
  and\ \bibinfo {author} {\bibfnamefont {M.~E.}\ \bibnamefont {Cates}},\
  }\bibfield  {title} {\bibinfo {title} {Continuum theory of phase separation
  kinetics for active brownian particles},\ }\href
  {https://doi.org/10.1103/PhysRevLett.111.145702} {\bibfield  {journal}
  {\bibinfo  {journal} {Phys. Rev. Lett.}\ }\textbf {\bibinfo {volume} {111}},\
  \bibinfo {pages} {145702} (\bibinfo {year} {2013})}\BibitemShut {NoStop}%
\bibitem [{\citenamefont {Bialk{\'{e}}}\ \emph {et~al.}(2013)\citenamefont
  {Bialk{\'{e}}}, \citenamefont {L{\"o}wen},\ and\ \citenamefont
  {Speck}}]{Bialk2013}%
  \BibitemOpen
  \bibfield  {author} {\bibinfo {author} {\bibfnamefont {J.}~\bibnamefont
  {Bialk{\'{e}}}}, \bibinfo {author} {\bibfnamefont {H.}~\bibnamefont
  {L{\"o}wen}},\ and\ \bibinfo {author} {\bibfnamefont {T.}~\bibnamefont
  {Speck}},\ }\bibfield  {title} {\bibinfo {title} {Microscopic theory for the
  phase separation of self-propelled repulsive disks},\ }\href
  {https://doi.org/10.1209/0295-5075/103/30008} {\bibfield  {journal} {\bibinfo
   {journal} {{EPL} (Europhysics Letters)}\ }\textbf {\bibinfo {volume}
  {103}},\ \bibinfo {pages} {30008} (\bibinfo {year} {2013})}\BibitemShut
  {NoStop}%
\bibitem [{\citenamefont {Speck}\ \emph {et~al.}(2014)\citenamefont {Speck},
  \citenamefont {Bialk\'e}, \citenamefont {Menzel},\ and\ \citenamefont
  {L\"owen}}]{Speck2014PRL}%
  \BibitemOpen
  \bibfield  {author} {\bibinfo {author} {\bibfnamefont {T.}~\bibnamefont
  {Speck}}, \bibinfo {author} {\bibfnamefont {J.}~\bibnamefont {Bialk\'e}},
  \bibinfo {author} {\bibfnamefont {A.~M.}\ \bibnamefont {Menzel}},\ and\
  \bibinfo {author} {\bibfnamefont {H.}~\bibnamefont {L\"owen}},\ }\bibfield
  {title} {\bibinfo {title} {Effective cahn-hilliard equation for the phase
  separation of active brownian particles},\ }\href
  {https://doi.org/10.1103/PhysRevLett.112.218304} {\bibfield  {journal}
  {\bibinfo  {journal} {Phys. Rev. Lett.}\ }\textbf {\bibinfo {volume} {112}},\
  \bibinfo {pages} {218304} (\bibinfo {year} {2014})}\BibitemShut {NoStop}%
\bibitem [{\citenamefont {Speck}\ \emph {et~al.}(2015)\citenamefont {Speck},
  \citenamefont {Menzel}, \citenamefont {Bialk{\'e}},\ and\ \citenamefont
  {L{\"o}wen}}]{Speck2015}%
  \BibitemOpen
  \bibfield  {author} {\bibinfo {author} {\bibfnamefont {T.}~\bibnamefont
  {Speck}}, \bibinfo {author} {\bibfnamefont {A.~M.}\ \bibnamefont {Menzel}},
  \bibinfo {author} {\bibfnamefont {J.}~\bibnamefont {Bialk{\'e}}},\ and\
  \bibinfo {author} {\bibfnamefont {H.}~\bibnamefont {L{\"o}wen}},\ }\bibfield
  {title} {\bibinfo {title} {Dynamical mean-field theory and weakly non-linear
  analysis for the phase separation of active brownian particles},\ }\href
  {https://doi.org/10.1063/1.4922324} {\bibfield  {journal} {\bibinfo
  {journal} {The Journal of Chemical Physics}\ }\textbf {\bibinfo {volume}
  {142}},\ \bibinfo {pages} {224109} (\bibinfo {year} {2015})}\BibitemShut
  {NoStop}%
\bibitem [{\citenamefont {{Wittkowski}}\ \emph {et~al.}(2014)\citenamefont
  {{Wittkowski}}, \citenamefont {{Tiribocchi}}, \citenamefont {{Stenhammar}},
  \citenamefont {{Allen}}, \citenamefont {{Marenduzzo}},\ and\ \citenamefont
  {{Cates}}}]{Wittkowski2014NatCom}%
  \BibitemOpen
  \bibfield  {author} {\bibinfo {author} {\bibfnamefont {R.}~\bibnamefont
  {{Wittkowski}}}, \bibinfo {author} {\bibfnamefont {A.}~\bibnamefont
  {{Tiribocchi}}}, \bibinfo {author} {\bibfnamefont {J.}~\bibnamefont
  {{Stenhammar}}}, \bibinfo {author} {\bibfnamefont {R.~J.}\ \bibnamefont
  {{Allen}}}, \bibinfo {author} {\bibfnamefont {D.}~\bibnamefont
  {{Marenduzzo}}},\ and\ \bibinfo {author} {\bibfnamefont {M.~E.}\ \bibnamefont
  {{Cates}}},\ }\bibfield  {title} {\bibinfo {title} {{Scalar
  {\ensuremath{\varphi}}$^{4}$ field theory for active-particle phase
  separation}},\ }\href {https://doi.org/10.1038/ncomms5351} {\bibfield
  {journal} {\bibinfo  {journal} {Nature Communications}\ }\textbf {\bibinfo
  {volume} {5}},\ \bibinfo {eid} {4351} (\bibinfo {year} {2014})}\BibitemShut
  {NoStop}%
\bibitem [{\citenamefont {Cates}\ and\ \citenamefont
  {Tailleur}(2015)}]{Cates2015Annual_Review}%
  \BibitemOpen
  \bibfield  {author} {\bibinfo {author} {\bibfnamefont {M.~E.}\ \bibnamefont
  {Cates}}\ and\ \bibinfo {author} {\bibfnamefont {J.}~\bibnamefont
  {Tailleur}},\ }\bibfield  {title} {\bibinfo {title} {Motility-induced phase
  separation},\ }\href
  {https://doi.org/10.1146/annurev-conmatphys-031214-014710} {\bibfield
  {journal} {\bibinfo  {journal} {Annual Review of Condensed Matter Physics}\
  }\textbf {\bibinfo {volume} {6}},\ \bibinfo {pages} {219} (\bibinfo {year}
  {2015})}\BibitemShut {NoStop}%
\bibitem [{\citenamefont {Redner}\ \emph {et~al.}(2016)\citenamefont {Redner},
  \citenamefont {Wagner}, \citenamefont {Baskaran},\ and\ \citenamefont
  {Hagan}}]{RednerPRL2016}%
  \BibitemOpen
  \bibfield  {author} {\bibinfo {author} {\bibfnamefont {G.~S.}\ \bibnamefont
  {Redner}}, \bibinfo {author} {\bibfnamefont {C.~G.}\ \bibnamefont {Wagner}},
  \bibinfo {author} {\bibfnamefont {A.}~\bibnamefont {Baskaran}},\ and\
  \bibinfo {author} {\bibfnamefont {M.~F.}\ \bibnamefont {Hagan}},\ }\bibfield
  {title} {\bibinfo {title} {Classical nucleation theory description of active
  colloid assembly},\ }\href {https://doi.org/10.1103/PhysRevLett.117.148002}
  {\bibfield  {journal} {\bibinfo  {journal} {Phys. Rev. Lett.}\ }\textbf
  {\bibinfo {volume} {117}},\ \bibinfo {pages} {148002} (\bibinfo {year}
  {2016})}\BibitemShut {NoStop}%
\bibitem [{\citenamefont {Solon}\ \emph
  {et~al.}(2018{\natexlab{a}})\citenamefont {Solon}, \citenamefont
  {Stenhammar}, \citenamefont {Cates}, \citenamefont {Kafri},\ and\
  \citenamefont {Tailleur}}]{Solon2018PRE}%
  \BibitemOpen
  \bibfield  {author} {\bibinfo {author} {\bibfnamefont {A.~P.}\ \bibnamefont
  {Solon}}, \bibinfo {author} {\bibfnamefont {J.}~\bibnamefont {Stenhammar}},
  \bibinfo {author} {\bibfnamefont {M.~E.}\ \bibnamefont {Cates}}, \bibinfo
  {author} {\bibfnamefont {Y.}~\bibnamefont {Kafri}},\ and\ \bibinfo {author}
  {\bibfnamefont {J.}~\bibnamefont {Tailleur}},\ }\bibfield  {title} {\bibinfo
  {title} {Generalized thermodynamics of phase equilibria in scalar active
  matter},\ }\href {https://doi.org/10.1103/PhysRevE.97.020602} {\bibfield
  {journal} {\bibinfo  {journal} {Phys. Rev. E}\ }\textbf {\bibinfo {volume}
  {97}},\ \bibinfo {pages} {020602(R)} (\bibinfo {year}
  {2018}{\natexlab{a}})}\BibitemShut {NoStop}%
\bibitem [{\citenamefont {Arnoulx~de Pirey}\ \emph {et~al.}(2019)\citenamefont
  {Arnoulx~de Pirey}, \citenamefont {Lozano},\ and\ \citenamefont {van
  Wijland}}]{dePirey2019PRL}%
  \BibitemOpen
  \bibfield  {author} {\bibinfo {author} {\bibfnamefont {T.}~\bibnamefont
  {Arnoulx~de Pirey}}, \bibinfo {author} {\bibfnamefont {G.}~\bibnamefont
  {Lozano}},\ and\ \bibinfo {author} {\bibfnamefont {F.}~\bibnamefont {van
  Wijland}},\ }\bibfield  {title} {\bibinfo {title} {Active hard spheres in
  infinitely many dimensions},\ }\href
  {https://doi.org/10.1103/PhysRevLett.123.260602} {\bibfield  {journal}
  {\bibinfo  {journal} {Phys. Rev. Lett.}\ }\textbf {\bibinfo {volume} {123}},\
  \bibinfo {pages} {260602} (\bibinfo {year} {2019})}\BibitemShut {NoStop}%
\bibitem [{\citenamefont {Redner}\ \emph {et~al.}(2013)\citenamefont {Redner},
  \citenamefont {Hagan},\ and\ \citenamefont {Baskaran}}]{Redner2013PRL}%
  \BibitemOpen
  \bibfield  {author} {\bibinfo {author} {\bibfnamefont {G.~S.}\ \bibnamefont
  {Redner}}, \bibinfo {author} {\bibfnamefont {M.~F.}\ \bibnamefont {Hagan}},\
  and\ \bibinfo {author} {\bibfnamefont {A.}~\bibnamefont {Baskaran}},\
  }\bibfield  {title} {\bibinfo {title} {Structure and dynamics of a
  phase-separating active colloidal fluid},\ }\href
  {https://doi.org/10.1103/PhysRevLett.110.055701} {\bibfield  {journal}
  {\bibinfo  {journal} {Phys. Rev. Lett.}\ }\textbf {\bibinfo {volume} {110}},\
  \bibinfo {pages} {055701} (\bibinfo {year} {2013})}\BibitemShut {NoStop}%
\bibitem [{\citenamefont {{Fily}}\ \emph {et~al.}(2014)\citenamefont {{Fily}},
  \citenamefont {{Henkes}},\ and\ \citenamefont
  {{Marchetti}}}]{Fily2014soft_matter}%
  \BibitemOpen
  \bibfield  {author} {\bibinfo {author} {\bibfnamefont {Y.}~\bibnamefont
  {{Fily}}}, \bibinfo {author} {\bibfnamefont {S.}~\bibnamefont {{Henkes}}},\
  and\ \bibinfo {author} {\bibfnamefont {M.~C.}\ \bibnamefont {{Marchetti}}},\
  }\bibfield  {title} {\bibinfo {title} {{Freezing and phase separation of
  self-propelled disks}},\ }\href {https://doi.org/10.1039/c3sm52469h}
  {\bibfield  {journal} {\bibinfo  {journal} {Soft Matter}\ }\textbf {\bibinfo
  {volume} {10}},\ \bibinfo {pages} {2132} (\bibinfo {year}
  {2014})}\BibitemShut {NoStop}%
\bibitem [{\citenamefont {{Stenhammar}}\ \emph {et~al.}(2014)\citenamefont
  {{Stenhammar}}, \citenamefont {{Marenduzzo}}, \citenamefont {{Allen}},\ and\
  \citenamefont {{Cates}}}]{Stenhammar2014soft_matter}%
  \BibitemOpen
  \bibfield  {author} {\bibinfo {author} {\bibfnamefont {J.}~\bibnamefont
  {{Stenhammar}}}, \bibinfo {author} {\bibfnamefont {D.}~\bibnamefont
  {{Marenduzzo}}}, \bibinfo {author} {\bibfnamefont {R.~J.}\ \bibnamefont
  {{Allen}}},\ and\ \bibinfo {author} {\bibfnamefont {M.~E.}\ \bibnamefont
  {{Cates}}},\ }\bibfield  {title} {\bibinfo {title} {{Phase behaviour of
  active Brownian particles: the role of dimensionality}},\ }\href
  {https://doi.org/10.1039/c3sm52813h} {\bibfield  {journal} {\bibinfo
  {journal} {Soft Matter}\ }\textbf {\bibinfo {volume} {10}},\ \bibinfo {pages}
  {1489} (\bibinfo {year} {2014})}\BibitemShut {NoStop}%
\bibitem [{\citenamefont {Levis}\ \emph {et~al.}(2017)\citenamefont {Levis},
  \citenamefont {Codina},\ and\ \citenamefont
  {Pagonabarraga}}]{Levis2017SoftMatter}%
  \BibitemOpen
  \bibfield  {author} {\bibinfo {author} {\bibfnamefont {D.}~\bibnamefont
  {Levis}}, \bibinfo {author} {\bibfnamefont {J.}~\bibnamefont {Codina}},\ and\
  \bibinfo {author} {\bibfnamefont {I.}~\bibnamefont {Pagonabarraga}},\
  }\bibfield  {title} {\bibinfo {title} {Active brownian equation of state:
  metastability and phase coexistence},\ }\href
  {https://doi.org/10.1039/C7SM01504F} {\bibfield  {journal} {\bibinfo
  {journal} {Soft Matter}\ }\textbf {\bibinfo {volume} {13}},\ \bibinfo {pages}
  {8113} (\bibinfo {year} {2017})}\BibitemShut {NoStop}%
\bibitem [{\citenamefont {Siebert}\ \emph {et~al.}(2018)\citenamefont
  {Siebert}, \citenamefont {Dittrich}, \citenamefont {Schmid}, \citenamefont
  {Binder}, \citenamefont {Speck},\ and\ \citenamefont
  {Virnau}}]{Siebert2018PRE}%
  \BibitemOpen
  \bibfield  {author} {\bibinfo {author} {\bibfnamefont {J.~T.}\ \bibnamefont
  {Siebert}}, \bibinfo {author} {\bibfnamefont {F.}~\bibnamefont {Dittrich}},
  \bibinfo {author} {\bibfnamefont {F.}~\bibnamefont {Schmid}}, \bibinfo
  {author} {\bibfnamefont {K.}~\bibnamefont {Binder}}, \bibinfo {author}
  {\bibfnamefont {T.}~\bibnamefont {Speck}},\ and\ \bibinfo {author}
  {\bibfnamefont {P.}~\bibnamefont {Virnau}},\ }\bibfield  {title} {\bibinfo
  {title} {Critical behavior of active brownian particles},\ }\href
  {https://doi.org/10.1103/PhysRevE.98.030601} {\bibfield  {journal} {\bibinfo
  {journal} {Phys. Rev. E}\ }\textbf {\bibinfo {volume} {98}},\ \bibinfo
  {pages} {030601(R)} (\bibinfo {year} {2018})}\BibitemShut {NoStop}%
\bibitem [{\citenamefont {Digregorio}\ \emph {et~al.}(2018)\citenamefont
  {Digregorio}, \citenamefont {Levis}, \citenamefont {Suma}, \citenamefont
  {Cugliandolo}, \citenamefont {Gonnella},\ and\ \citenamefont
  {Pagonabarraga}}]{Digregorio2018PRL}%
  \BibitemOpen
  \bibfield  {author} {\bibinfo {author} {\bibfnamefont {P.}~\bibnamefont
  {Digregorio}}, \bibinfo {author} {\bibfnamefont {D.}~\bibnamefont {Levis}},
  \bibinfo {author} {\bibfnamefont {A.}~\bibnamefont {Suma}}, \bibinfo {author}
  {\bibfnamefont {L.~F.}\ \bibnamefont {Cugliandolo}}, \bibinfo {author}
  {\bibfnamefont {G.}~\bibnamefont {Gonnella}},\ and\ \bibinfo {author}
  {\bibfnamefont {I.}~\bibnamefont {Pagonabarraga}},\ }\bibfield  {title}
  {\bibinfo {title} {Full phase diagram of active brownian disks: From melting
  to motility-induced phase separation},\ }\href
  {https://doi.org/10.1103/PhysRevLett.121.098003} {\bibfield  {journal}
  {\bibinfo  {journal} {Phys. Rev. Lett.}\ }\textbf {\bibinfo {volume} {121}},\
  \bibinfo {pages} {098003} (\bibinfo {year} {2018})}\BibitemShut {NoStop}%
\bibitem [{\citenamefont {Caporusso}\ \emph {et~al.}(2020)\citenamefont
  {Caporusso}, \citenamefont {Digregorio}, \citenamefont {Levis}, \citenamefont
  {Cugliandolo},\ and\ \citenamefont {Gonnella}}]{Caporusso2020PRL}%
  \BibitemOpen
  \bibfield  {author} {\bibinfo {author} {\bibfnamefont {C.~B.}\ \bibnamefont
  {Caporusso}}, \bibinfo {author} {\bibfnamefont {P.}~\bibnamefont
  {Digregorio}}, \bibinfo {author} {\bibfnamefont {D.}~\bibnamefont {Levis}},
  \bibinfo {author} {\bibfnamefont {L.~F.}\ \bibnamefont {Cugliandolo}},\ and\
  \bibinfo {author} {\bibfnamefont {G.}~\bibnamefont {Gonnella}},\ }\bibfield
  {title} {\bibinfo {title} {Motility-induced microphase and macrophase
  separation in a two-dimensional active brownian particle system},\ }\href
  {https://doi.org/10.1103/PhysRevLett.125.178004} {\bibfield  {journal}
  {\bibinfo  {journal} {Phys. Rev. Lett.}\ }\textbf {\bibinfo {volume} {125}},\
  \bibinfo {pages} {178004} (\bibinfo {year} {2020})}\BibitemShut {NoStop}%
\bibitem [{\citenamefont {Farage}\ \emph {et~al.}(2015)\citenamefont {Farage},
  \citenamefont {Krinninger},\ and\ \citenamefont {Brader}}]{Farage2015PRE}%
  \BibitemOpen
  \bibfield  {author} {\bibinfo {author} {\bibfnamefont {T.~F.~F.}\
  \bibnamefont {Farage}}, \bibinfo {author} {\bibfnamefont {P.}~\bibnamefont
  {Krinninger}},\ and\ \bibinfo {author} {\bibfnamefont {J.~M.}\ \bibnamefont
  {Brader}},\ }\bibfield  {title} {\bibinfo {title} {Effective interactions in
  active brownian suspensions},\ }\href
  {https://doi.org/10.1103/PhysRevE.91.042310} {\bibfield  {journal} {\bibinfo
  {journal} {Phys. Rev. E}\ }\textbf {\bibinfo {volume} {91}},\ \bibinfo
  {pages} {042310} (\bibinfo {year} {2015})}\BibitemShut {NoStop}%
\bibitem [{\citenamefont {Solon}\ \emph
  {et~al.}(2018{\natexlab{b}})\citenamefont {Solon}, \citenamefont
  {Stenhammar}, \citenamefont {Cates}, \citenamefont {Kafri},\ and\
  \citenamefont {Tailleur}}]{Solon_2018}%
  \BibitemOpen
  \bibfield  {author} {\bibinfo {author} {\bibfnamefont {A.~P.}\ \bibnamefont
  {Solon}}, \bibinfo {author} {\bibfnamefont {J.}~\bibnamefont {Stenhammar}},
  \bibinfo {author} {\bibfnamefont {M.~E.}\ \bibnamefont {Cates}}, \bibinfo
  {author} {\bibfnamefont {Y.}~\bibnamefont {Kafri}},\ and\ \bibinfo {author}
  {\bibfnamefont {J.}~\bibnamefont {Tailleur}},\ }\bibfield  {title} {\bibinfo
  {title} {Generalized thermodynamics of motility-induced phase separation:
  phase equilibria, laplace pressure, and change of ensembles},\ }\href
  {https://doi.org/10.1088/1367-2630/aaccdd} {\bibfield  {journal} {\bibinfo
  {journal} {New Journal of Physics}\ }\textbf {\bibinfo {volume} {20}},\
  \bibinfo {pages} {075001} (\bibinfo {year} {2018}{\natexlab{b}})}\BibitemShut
  {NoStop}%
\bibitem [{\citenamefont {Speck}(2021)}]{Speck2021pre}%
  \BibitemOpen
  \bibfield  {author} {\bibinfo {author} {\bibfnamefont {T.}~\bibnamefont
  {Speck}},\ }\bibfield  {title} {\bibinfo {title} {Coexistence of active
  brownian disks: van der waals theory and analytical results},\ }\href
  {https://doi.org/10.1103/PhysRevE.103.012607} {\bibfield  {journal} {\bibinfo
   {journal} {Phys. Rev. E}\ }\textbf {\bibinfo {volume} {103}},\ \bibinfo
  {pages} {012607} (\bibinfo {year} {2021})}\BibitemShut {NoStop}%
\bibitem [{\citenamefont {Bialk\'e}\ \emph {et~al.}(2015)\citenamefont
  {Bialk\'e}, \citenamefont {Siebert}, \citenamefont {L\"owen},\ and\
  \citenamefont {Speck}}]{Bialke2015PRL}%
  \BibitemOpen
  \bibfield  {author} {\bibinfo {author} {\bibfnamefont {J.}~\bibnamefont
  {Bialk\'e}}, \bibinfo {author} {\bibfnamefont {J.~T.}\ \bibnamefont
  {Siebert}}, \bibinfo {author} {\bibfnamefont {H.}~\bibnamefont {L\"owen}},\
  and\ \bibinfo {author} {\bibfnamefont {T.}~\bibnamefont {Speck}},\ }\bibfield
   {title} {\bibinfo {title} {Negative interfacial tension in phase-separated
  active brownian particles},\ }\href
  {https://doi.org/10.1103/PhysRevLett.115.098301} {\bibfield  {journal}
  {\bibinfo  {journal} {Phys. Rev. Lett.}\ }\textbf {\bibinfo {volume} {115}},\
  \bibinfo {pages} {098301} (\bibinfo {year} {2015})}\BibitemShut {NoStop}%
\bibitem [{\citenamefont {Tjhung}\ \emph {et~al.}(2018)\citenamefont {Tjhung},
  \citenamefont {Nardini},\ and\ \citenamefont {Cates}}]{Tjhung2018PRX}%
  \BibitemOpen
  \bibfield  {author} {\bibinfo {author} {\bibfnamefont {E.}~\bibnamefont
  {Tjhung}}, \bibinfo {author} {\bibfnamefont {C.}~\bibnamefont {Nardini}},\
  and\ \bibinfo {author} {\bibfnamefont {M.~E.}\ \bibnamefont {Cates}},\
  }\bibfield  {title} {\bibinfo {title} {Cluster phases and bubbly phase
  separation in active fluids: Reversal of the ostwald process},\ }\href
  {https://doi.org/10.1103/PhysRevX.8.031080} {\bibfield  {journal} {\bibinfo
  {journal} {Phys. Rev. X}\ }\textbf {\bibinfo {volume} {8}},\ \bibinfo {pages}
  {031080} (\bibinfo {year} {2018})}\BibitemShut {NoStop}%
\bibitem [{\citenamefont {Shi}\ \emph {et~al.}(2020)\citenamefont {Shi},
  \citenamefont {Fausti}, \citenamefont {Chat\'e}, \citenamefont {Nardini},\
  and\ \citenamefont {Solon}}]{Shi2020PRL}%
  \BibitemOpen
  \bibfield  {author} {\bibinfo {author} {\bibfnamefont {X.-q.}\ \bibnamefont
  {Shi}}, \bibinfo {author} {\bibfnamefont {G.}~\bibnamefont {Fausti}},
  \bibinfo {author} {\bibfnamefont {H.}~\bibnamefont {Chat\'e}}, \bibinfo
  {author} {\bibfnamefont {C.}~\bibnamefont {Nardini}},\ and\ \bibinfo {author}
  {\bibfnamefont {A.}~\bibnamefont {Solon}},\ }\bibfield  {title} {\bibinfo
  {title} {Self-organized critical coexistence phase in repulsive active
  particles},\ }\href {https://doi.org/10.1103/PhysRevLett.125.168001}
  {\bibfield  {journal} {\bibinfo  {journal} {Phys. Rev. Lett.}\ }\textbf
  {\bibinfo {volume} {125}},\ \bibinfo {pages} {168001} (\bibinfo {year}
  {2020})}\BibitemShut {NoStop}%
\bibitem [{\citenamefont {Caprini}\ \emph
  {et~al.}(2020{\natexlab{a}})\citenamefont {Caprini}, \citenamefont {Marini
  Bettolo~Marconi},\ and\ \citenamefont {Puglisi}}]{Caprini2020PRL}%
  \BibitemOpen
  \bibfield  {author} {\bibinfo {author} {\bibfnamefont {L.}~\bibnamefont
  {Caprini}}, \bibinfo {author} {\bibfnamefont {U.}~\bibnamefont {Marini
  Bettolo~Marconi}},\ and\ \bibinfo {author} {\bibfnamefont {A.}~\bibnamefont
  {Puglisi}},\ }\bibfield  {title} {\bibinfo {title} {Spontaneous velocity
  alignment in motility-induced phase separation},\ }\href
  {https://doi.org/10.1103/PhysRevLett.124.078001} {\bibfield  {journal}
  {\bibinfo  {journal} {Phys. Rev. Lett.}\ }\textbf {\bibinfo {volume} {124}},\
  \bibinfo {pages} {078001} (\bibinfo {year} {2020}{\natexlab{a}})}\BibitemShut
  {NoStop}%
\bibitem [{\citenamefont {Caprini}\ \emph
  {et~al.}(2020{\natexlab{b}})\citenamefont {Caprini}, \citenamefont {Marconi},
  \citenamefont {Maggi}, \citenamefont {Paoluzzi},\ and\ \citenamefont
  {Puglisi}}]{Caprini2020PRR}%
  \BibitemOpen
  \bibfield  {author} {\bibinfo {author} {\bibfnamefont {L.}~\bibnamefont
  {Caprini}}, \bibinfo {author} {\bibfnamefont {U.~M.~B.}\ \bibnamefont
  {Marconi}}, \bibinfo {author} {\bibfnamefont {C.}~\bibnamefont {Maggi}},
  \bibinfo {author} {\bibfnamefont {M.}~\bibnamefont {Paoluzzi}},\ and\
  \bibinfo {author} {\bibfnamefont {A.}~\bibnamefont {Puglisi}},\ }\bibfield
  {title} {\bibinfo {title} {Hidden velocity ordering in dense suspensions of
  self-propelled disks},\ }\href
  {https://doi.org/10.1103/PhysRevResearch.2.023321} {\bibfield  {journal}
  {\bibinfo  {journal} {Phys. Rev. Research}\ }\textbf {\bibinfo {volume}
  {2}},\ \bibinfo {pages} {023321} (\bibinfo {year}
  {2020}{\natexlab{b}})}\BibitemShut {NoStop}%
\bibitem [{\citenamefont {{Caprini}}\ and\ \citenamefont {{Marini Bettolo
  Marconi}}(2021)}]{Caprini2021Soft_Matter}%
  \BibitemOpen
  \bibfield  {author} {\bibinfo {author} {\bibfnamefont {L.}~\bibnamefont
  {{Caprini}}}\ and\ \bibinfo {author} {\bibfnamefont {U.}~\bibnamefont
  {{Marini Bettolo Marconi}}},\ }\bibfield  {title} {\bibinfo {title} {{Spatial
  velocity correlations in inertial systems of active Brownian particles}},\
  }\href {https://doi.org/10.1039/D0SM02273J} {\bibfield  {journal} {\bibinfo
  {journal} {Soft Matter}\ }\textbf {\bibinfo {volume} {17}},\ \bibinfo {pages}
  {4109} (\bibinfo {year} {2021})}\BibitemShut {NoStop}%
\bibitem [{\citenamefont {Flenner}\ \emph {et~al.}(2016)\citenamefont
  {Flenner}, \citenamefont {Szamel},\ and\ \citenamefont
  {Berthier}}]{Flenner2016SoftMatter}%
  \BibitemOpen
  \bibfield  {author} {\bibinfo {author} {\bibfnamefont {E.}~\bibnamefont
  {Flenner}}, \bibinfo {author} {\bibfnamefont {G.}~\bibnamefont {Szamel}},\
  and\ \bibinfo {author} {\bibfnamefont {L.}~\bibnamefont {Berthier}},\
  }\bibfield  {title} {\bibinfo {title} {The nonequilibrium glassy dynamics of
  self-propelled particles},\ }\href {https://doi.org/10.1039/C6SM01322H}
  {\bibfield  {journal} {\bibinfo  {journal} {Soft Matter}\ }\textbf {\bibinfo
  {volume} {12}},\ \bibinfo {pages} {7136} (\bibinfo {year}
  {2016})}\BibitemShut {NoStop}%
\bibitem [{\citenamefont {{Henkes}}\ \emph {et~al.}(2020)\citenamefont
  {{Henkes}}, \citenamefont {{Kostanjevec}}, \citenamefont {{Collinson}},
  \citenamefont {{Sknepnek}},\ and\ \citenamefont
  {{Bertin}}}]{Henkes2020Nature_Communications}%
  \BibitemOpen
  \bibfield  {author} {\bibinfo {author} {\bibfnamefont {S.}~\bibnamefont
  {{Henkes}}}, \bibinfo {author} {\bibfnamefont {K.}~\bibnamefont
  {{Kostanjevec}}}, \bibinfo {author} {\bibfnamefont {J.~M.}\ \bibnamefont
  {{Collinson}}}, \bibinfo {author} {\bibfnamefont {R.}~\bibnamefont
  {{Sknepnek}}},\ and\ \bibinfo {author} {\bibfnamefont {E.}~\bibnamefont
  {{Bertin}}},\ }\bibfield  {title} {\bibinfo {title} {{Dense active matter
  model of motion patterns in confluent cell monolayers}},\ }\href
  {https://doi.org/10.1038/s41467-020-15164-5} {\bibfield  {journal} {\bibinfo
  {journal} {Nature Communications}\ }\textbf {\bibinfo {volume} {11}},\
  \bibinfo {eid} {1405} (\bibinfo {year} {2020})}\BibitemShut {NoStop}%
\bibitem [{\citenamefont {Szamel}\ and\ \citenamefont
  {Flenner}(2021)}]{Szamel2021EPL}%
  \BibitemOpen
  \bibfield  {author} {\bibinfo {author} {\bibfnamefont {G.}~\bibnamefont
  {Szamel}}\ and\ \bibinfo {author} {\bibfnamefont {E.}~\bibnamefont
  {Flenner}},\ }\bibfield  {title} {\bibinfo {title} {Long-ranged velocity
  correlations in dense systems of self-propelled particles},\ }\href
  {https://doi.org/10.1209/0295-5075/133/60002} {\bibfield  {journal} {\bibinfo
   {journal} {{EPL} (Europhysics Letters)}\ }\textbf {\bibinfo {volume}
  {133}},\ \bibinfo {pages} {60002} (\bibinfo {year} {2021})}\BibitemShut
  {NoStop}%
\bibitem [{\citenamefont {Keta}\ \emph {et~al.}(2022)\citenamefont {Keta},
  \citenamefont {Jack},\ and\ \citenamefont {Berthier}}]{Keta2022}%
  \BibitemOpen
  \bibfield  {author} {\bibinfo {author} {\bibfnamefont {Y.-E.}\ \bibnamefont
  {Keta}}, \bibinfo {author} {\bibfnamefont {R.~L.}\ \bibnamefont {Jack}},\
  and\ \bibinfo {author} {\bibfnamefont {L.}~\bibnamefont {Berthier}},\
  }\bibfield  {title} {\bibinfo {title} {Disordered collective motion in dense
  assemblies of persistent particles},\ }\href
  {https://doi.org/10.1103/PhysRevLett.129.048002} {\bibfield  {journal}
  {\bibinfo  {journal} {Phys. Rev. Lett.}\ }\textbf {\bibinfo {volume} {129}},\
  \bibinfo {pages} {048002} (\bibinfo {year} {2022})}\BibitemShut {NoStop}%
\bibitem [{\citenamefont {Toner}\ and\ \citenamefont
  {Tu}(1995)}]{TonerPRL1995}%
  \BibitemOpen
  \bibfield  {author} {\bibinfo {author} {\bibfnamefont {J.}~\bibnamefont
  {Toner}}\ and\ \bibinfo {author} {\bibfnamefont {Y.}~\bibnamefont {Tu}},\
  }\bibfield  {title} {\bibinfo {title} {Long-range order in a two-dimensional
  dynamical $\mathrm{XY}$ model: How birds fly together},\ }\href
  {https://doi.org/10.1103/PhysRevLett.75.4326} {\bibfield  {journal} {\bibinfo
   {journal} {Phys. Rev. Lett.}\ }\textbf {\bibinfo {volume} {75}},\ \bibinfo
  {pages} {4326} (\bibinfo {year} {1995})}\BibitemShut {NoStop}%
\bibitem [{\citenamefont {{Toner}}\ \emph {et~al.}(2005)\citenamefont
  {{Toner}}, \citenamefont {{Tu}},\ and\ \citenamefont
  {{Ramaswamy}}}]{Toner2005}%
  \BibitemOpen
  \bibfield  {author} {\bibinfo {author} {\bibfnamefont {J.}~\bibnamefont
  {{Toner}}}, \bibinfo {author} {\bibfnamefont {Y.}~\bibnamefont {{Tu}}},\ and\
  \bibinfo {author} {\bibfnamefont {S.}~\bibnamefont {{Ramaswamy}}},\
  }\bibfield  {title} {\bibinfo {title} {{Hydrodynamics and phases of
  flocks}},\ }\href {https://doi.org/10.1016/j.aop.2005.04.011} {\bibfield
  {journal} {\bibinfo  {journal} {Annals of Physics}\ }\textbf {\bibinfo
  {volume} {318}},\ \bibinfo {pages} {170} (\bibinfo {year}
  {2005})}\BibitemShut {NoStop}%
\bibitem [{\citenamefont {Chat\'e}\ \emph {et~al.}(2008)\citenamefont
  {Chat\'e}, \citenamefont {Ginelli}, \citenamefont {Gr\'egoire},\ and\
  \citenamefont {Raynaud}}]{Chate2008PRE}%
  \BibitemOpen
  \bibfield  {author} {\bibinfo {author} {\bibfnamefont {H.}~\bibnamefont
  {Chat\'e}}, \bibinfo {author} {\bibfnamefont {F.}~\bibnamefont {Ginelli}},
  \bibinfo {author} {\bibfnamefont {G.}~\bibnamefont {Gr\'egoire}},\ and\
  \bibinfo {author} {\bibfnamefont {F.}~\bibnamefont {Raynaud}},\ }\bibfield
  {title} {\bibinfo {title} {Collective motion of self-propelled particles
  interacting without cohesion},\ }\href
  {https://doi.org/10.1103/PhysRevE.77.046113} {\bibfield  {journal} {\bibinfo
  {journal} {Phys. Rev. E}\ }\textbf {\bibinfo {volume} {77}},\ \bibinfo
  {pages} {046113} (\bibinfo {year} {2008})}\BibitemShut {NoStop}%
\bibitem [{\citenamefont {Mandal}\ \emph {et~al.}(2019)\citenamefont {Mandal},
  \citenamefont {Liebchen},\ and\ \citenamefont {L\"owen}}]{Mandal2019PRL}%
  \BibitemOpen
  \bibfield  {author} {\bibinfo {author} {\bibfnamefont {S.}~\bibnamefont
  {Mandal}}, \bibinfo {author} {\bibfnamefont {B.}~\bibnamefont {Liebchen}},\
  and\ \bibinfo {author} {\bibfnamefont {H.}~\bibnamefont {L\"owen}},\
  }\bibfield  {title} {\bibinfo {title} {Motility-induced temperature
  difference in coexisting phases},\ }\href
  {https://doi.org/10.1103/PhysRevLett.123.228001} {\bibfield  {journal}
  {\bibinfo  {journal} {Phys. Rev. Lett.}\ }\textbf {\bibinfo {volume} {123}},\
  \bibinfo {pages} {228001} (\bibinfo {year} {2019})}\BibitemShut {NoStop}%
\bibitem [{\citenamefont {{Weeks}}\ \emph {et~al.}(1971)\citenamefont
  {{Weeks}}, \citenamefont {{Chandler}},\ and\ \citenamefont
  {{Andersen}}}]{WCA}%
  \BibitemOpen
  \bibfield  {author} {\bibinfo {author} {\bibfnamefont {J.~D.}\ \bibnamefont
  {{Weeks}}}, \bibinfo {author} {\bibfnamefont {D.}~\bibnamefont
  {{Chandler}}},\ and\ \bibinfo {author} {\bibfnamefont {H.~C.}\ \bibnamefont
  {{Andersen}}},\ }\bibfield  {title} {\bibinfo {title} {{Role of Repulsive
  Forces in Determining the Equilibrium Structure of Simple Liquids}},\ }\href
  {https://doi.org/10.1063/1.1674820} {\bibfield  {journal} {\bibinfo
  {journal} {\jcp}\ }\textbf {\bibinfo {volume} {54}},\ \bibinfo {pages} {5237}
  (\bibinfo {year} {1971})}\BibitemShut {NoStop}%
\bibitem [{\citenamefont {Marconi}\ \emph {et~al.}(2021)\citenamefont
  {Marconi}, \citenamefont {Caprini},\ and\ \citenamefont
  {Puglisi}}]{marconi2021}%
  \BibitemOpen
  \bibfield  {author} {\bibinfo {author} {\bibfnamefont {U.~M.~B.}\
  \bibnamefont {Marconi}}, \bibinfo {author} {\bibfnamefont {L.}~\bibnamefont
  {Caprini}},\ and\ \bibinfo {author} {\bibfnamefont {A.}~\bibnamefont
  {Puglisi}},\ }\bibfield  {title} {\bibinfo {title} {Hydrodynamics of simple
  active liquids: the emergence of velocity correlations},\ }\href
  {https://doi.org/10.1088/1367-2630/ac2b54} {\bibfield  {journal} {\bibinfo
  {journal} {New Journal of Physics}\ }\textbf {\bibinfo {volume} {23}},\
  \bibinfo {pages} {103024} (\bibinfo {year} {2021})}\BibitemShut {NoStop}%
\bibitem [{\citenamefont {Wensink}\ and\ \citenamefont
  {L{\"o}wen}(2012)}]{Wensink_2012}%
  \BibitemOpen
  \bibfield  {author} {\bibinfo {author} {\bibfnamefont {H.~H.}\ \bibnamefont
  {Wensink}}\ and\ \bibinfo {author} {\bibfnamefont {H.}~\bibnamefont
  {L{\"o}wen}},\ }\bibfield  {title} {\bibinfo {title} {Emergent states in
  dense systems of active rods: from swarming to turbulence},\ }\href
  {https://doi.org/10.1088/0953-8984/24/46/464130} {\bibfield  {journal}
  {\bibinfo  {journal} {Journal of Physics: Condensed Matter}\ }\textbf
  {\bibinfo {volume} {24}},\ \bibinfo {pages} {464130} (\bibinfo {year}
  {2012})}\BibitemShut {NoStop}%
\bibitem [{\citenamefont {Dunkel}\ \emph {et~al.}(2013)\citenamefont {Dunkel},
  \citenamefont {Heidenreich}, \citenamefont {Drescher}, \citenamefont
  {Wensink}, \citenamefont {B\"ar},\ and\ \citenamefont
  {Goldstein}}]{Dunkel2013PRL}%
  \BibitemOpen
  \bibfield  {author} {\bibinfo {author} {\bibfnamefont {J.}~\bibnamefont
  {Dunkel}}, \bibinfo {author} {\bibfnamefont {S.}~\bibnamefont {Heidenreich}},
  \bibinfo {author} {\bibfnamefont {K.}~\bibnamefont {Drescher}}, \bibinfo
  {author} {\bibfnamefont {H.~H.}\ \bibnamefont {Wensink}}, \bibinfo {author}
  {\bibfnamefont {M.}~\bibnamefont {B\"ar}},\ and\ \bibinfo {author}
  {\bibfnamefont {R.~E.}\ \bibnamefont {Goldstein}},\ }\bibfield  {title}
  {\bibinfo {title} {Fluid dynamics of bacterial turbulence},\ }\href
  {https://doi.org/10.1103/PhysRevLett.110.228102} {\bibfield  {journal}
  {\bibinfo  {journal} {Phys. Rev. Lett.}\ }\textbf {\bibinfo {volume} {110}},\
  \bibinfo {pages} {228102} (\bibinfo {year} {2013})}\BibitemShut {NoStop}%
\bibitem [{\citenamefont {Nishiguchi}\ and\ \citenamefont
  {Sano}(2015)}]{Nisiguchi2015PRE}%
  \BibitemOpen
  \bibfield  {author} {\bibinfo {author} {\bibfnamefont {D.}~\bibnamefont
  {Nishiguchi}}\ and\ \bibinfo {author} {\bibfnamefont {M.}~\bibnamefont
  {Sano}},\ }\bibfield  {title} {\bibinfo {title} {Mesoscopic turbulence and
  local order in janus particles self-propelling under an ac electric field},\
  }\href {https://doi.org/10.1103/PhysRevE.92.052309} {\bibfield  {journal}
  {\bibinfo  {journal} {Phys. Rev. E}\ }\textbf {\bibinfo {volume} {92}},\
  \bibinfo {pages} {052309} (\bibinfo {year} {2015})}\BibitemShut {NoStop}%
\bibitem [{\citenamefont {Creppy}\ \emph {et~al.}(2015)\citenamefont {Creppy},
  \citenamefont {Praud}, \citenamefont {Druart}, \citenamefont {Kohnke},\ and\
  \citenamefont {Plourabou\'e}}]{Creppy2015PRE}%
  \BibitemOpen
  \bibfield  {author} {\bibinfo {author} {\bibfnamefont {A.}~\bibnamefont
  {Creppy}}, \bibinfo {author} {\bibfnamefont {O.}~\bibnamefont {Praud}},
  \bibinfo {author} {\bibfnamefont {X.}~\bibnamefont {Druart}}, \bibinfo
  {author} {\bibfnamefont {P.~L.}\ \bibnamefont {Kohnke}},\ and\ \bibinfo
  {author} {\bibfnamefont {F.}~\bibnamefont {Plourabou\'e}},\ }\bibfield
  {title} {\bibinfo {title} {Turbulence of swarming sperm},\ }\href
  {https://doi.org/10.1103/PhysRevE.92.032722} {\bibfield  {journal} {\bibinfo
  {journal} {Phys. Rev. E}\ }\textbf {\bibinfo {volume} {92}},\ \bibinfo
  {pages} {032722} (\bibinfo {year} {2015})}\BibitemShut {NoStop}%
\bibitem [{\citenamefont {Guillamat}\ \emph {et~al.}(2017)\citenamefont
  {Guillamat}, \citenamefont {Ign{\'e}s-Mullol},\ and\ \citenamefont
  {Sagu{\'e}s}}]{Guillamat:2017}%
  \BibitemOpen
  \bibfield  {author} {\bibinfo {author} {\bibfnamefont {P.}~\bibnamefont
  {Guillamat}}, \bibinfo {author} {\bibfnamefont {J.}~\bibnamefont
  {Ign{\'e}s-Mullol}},\ and\ \bibinfo {author} {\bibfnamefont {F.}~\bibnamefont
  {Sagu{\'e}s}},\ }\bibfield  {title} {\bibinfo {title} {Taming active
  turbulence with patterned soft interfaces},\ }\href
  {https://doi.org/10.1038/s41467-017-00617-1} {\bibfield  {journal} {\bibinfo
  {journal} {Nature Communications}\ }\textbf {\bibinfo {volume} {8}},\
  \bibinfo {pages} {564} (\bibinfo {year} {2017})}\BibitemShut {NoStop}%
\bibitem [{\citenamefont {Lin}\ \emph {et~al.}(2021)\citenamefont {Lin},
  \citenamefont {Zhang}, \citenamefont {Bi}, \citenamefont {Li},\ and\
  \citenamefont {Feng}}]{Lin:2021tr}%
  \BibitemOpen
  \bibfield  {author} {\bibinfo {author} {\bibfnamefont {S.-Z.}\ \bibnamefont
  {Lin}}, \bibinfo {author} {\bibfnamefont {W.-Y.}\ \bibnamefont {Zhang}},
  \bibinfo {author} {\bibfnamefont {D.}~\bibnamefont {Bi}}, \bibinfo {author}
  {\bibfnamefont {B.}~\bibnamefont {Li}},\ and\ \bibinfo {author}
  {\bibfnamefont {X.-Q.}\ \bibnamefont {Feng}},\ }\bibfield  {title} {\bibinfo
  {title} {Energetics of mesoscale cell turbulence in two-dimensional
  monolayers},\ }\href {https://doi.org/10.1038/s42005-021-00530-6} {\bibfield
  {journal} {\bibinfo  {journal} {Communications Physics}\ }\textbf {\bibinfo
  {volume} {4}},\ \bibinfo {pages} {21} (\bibinfo {year} {2021})}\BibitemShut
  {NoStop}%
\bibitem [{\citenamefont {Liu}\ \emph {et~al.}(2021)\citenamefont {Liu},
  \citenamefont {Zeng}, \citenamefont {Ma},\ and\ \citenamefont
  {Cheng}}]{liu2020}%
  \BibitemOpen
  \bibfield  {author} {\bibinfo {author} {\bibfnamefont {Z.}~\bibnamefont
  {Liu}}, \bibinfo {author} {\bibfnamefont {W.}~\bibnamefont {Zeng}}, \bibinfo
  {author} {\bibfnamefont {X.}~\bibnamefont {Ma}},\ and\ \bibinfo {author}
  {\bibfnamefont {X.}~\bibnamefont {Cheng}},\ }\bibfield  {title} {\bibinfo
  {title} {Density fluctuations and energy spectra of 3d bacterial
  suspensions},\ }\href {https://doi.org/10.1039/D1SM01183A} {\bibfield
  {journal} {\bibinfo  {journal} {Soft Matter}\ }\textbf {\bibinfo {volume}
  {17}},\ \bibinfo {pages} {10806} (\bibinfo {year} {2021})}\BibitemShut
  {NoStop}%
\bibitem [{\citenamefont {Qi}\ \emph {et~al.}(2022)\citenamefont {Qi},
  \citenamefont {Westphal}, \citenamefont {Gompper},\ and\ \citenamefont
  {Winkler}}]{qi2021}%
  \BibitemOpen
  \bibfield  {author} {\bibinfo {author} {\bibfnamefont {K.}~\bibnamefont
  {Qi}}, \bibinfo {author} {\bibfnamefont {E.}~\bibnamefont {Westphal}},
  \bibinfo {author} {\bibfnamefont {G.}~\bibnamefont {Gompper}},\ and\ \bibinfo
  {author} {\bibfnamefont {R.~G.}\ \bibnamefont {Winkler}},\ }\bibfield
  {title} {\bibinfo {title} {Emergence of active turbulence in microswimmer
  suspensions due to active hydrodynamic stress and volume exclusion},\ }\href
  {https://doi.org/10.1038/s42005-022-00820-7} {\bibfield  {journal} {\bibinfo
  {journal} {Communications Physics}\ }\textbf {\bibinfo {volume} {5}},\
  \bibinfo {pages} {49} (\bibinfo {year} {2022})}\BibitemShut {NoStop}%
\bibitem [{\citenamefont {Alert}\ \emph {et~al.}(2022)\citenamefont {Alert},
  \citenamefont {Casademunt},\ and\ \citenamefont {Joanny}}]{alert2021}%
  \BibitemOpen
  \bibfield  {author} {\bibinfo {author} {\bibfnamefont {R.}~\bibnamefont
  {Alert}}, \bibinfo {author} {\bibfnamefont {J.}~\bibnamefont {Casademunt}},\
  and\ \bibinfo {author} {\bibfnamefont {J.-F.}\ \bibnamefont {Joanny}},\
  }\bibfield  {title} {\bibinfo {title} {Active turbulence},\ }\href
  {https://doi.org/10.1146/annurev-conmatphys-082321-035957} {\bibfield
  {journal} {\bibinfo  {journal} {Annual Review of Condensed Matter Physics}\
  }\textbf {\bibinfo {volume} {13}},\ \bibinfo {pages} {null} (\bibinfo {year}
  {2022})}\BibitemShut {NoStop}%
\bibitem [{\citenamefont {Onuki}(2007)}]{onuki2007}%
  \BibitemOpen
  \bibfield  {author} {\bibinfo {author} {\bibfnamefont {A.}~\bibnamefont
  {Onuki}},\ }\href {https://books.google.co.jp/books?id=uhOUPwAACAAJ} {\emph
  {\bibinfo {title} {Phase Transition Dynamics}}}\ (\bibinfo  {publisher}
  {Cambridge University Press},\ \bibinfo {year} {2007})\BibitemShut {NoStop}%
\bibitem [{\citenamefont {Bray}(2002)}]{Bray2002}%
  \BibitemOpen
  \bibfield  {author} {\bibinfo {author} {\bibfnamefont {A.~J.}\ \bibnamefont
  {Bray}},\ }\bibfield  {title} {\bibinfo {title} {Theory of phase-ordering
  kinetics},\ }\href {https://doi.org/10.1080/00018730110117433} {\bibfield
  {journal} {\bibinfo  {journal} {Advances in Physics}\ }\textbf {\bibinfo
  {volume} {51}},\ \bibinfo {pages} {481} (\bibinfo {year} {2002})}\BibitemShut
  {NoStop}%
\bibitem [{\citenamefont {{Zhang}}\ \emph {et~al.}(2010)\citenamefont
  {{Zhang}}, \citenamefont {{Be'er}}, \citenamefont {{Florin}},\ and\
  \citenamefont {{Swinney}}}]{Zhang2010PNAS}%
  \BibitemOpen
  \bibfield  {author} {\bibinfo {author} {\bibfnamefont {H.~P.}\ \bibnamefont
  {{Zhang}}}, \bibinfo {author} {\bibfnamefont {A.}~\bibnamefont {{Be'er}}},
  \bibinfo {author} {\bibfnamefont {E.~L.}\ \bibnamefont {{Florin}}},\ and\
  \bibinfo {author} {\bibfnamefont {H.~L.}\ \bibnamefont {{Swinney}}},\
  }\bibfield  {title} {\bibinfo {title} {{Collective motion and density
  fluctuations in bacterial colonies}},\ }\href
  {https://doi.org/10.1073/pnas.1001651107} {\bibfield  {journal} {\bibinfo
  {journal} {Proceedings of the National Academy of Science}\ }\textbf
  {\bibinfo {volume} {107}},\ \bibinfo {pages} {13626} (\bibinfo {year}
  {2010})}\BibitemShut {NoStop}%
\bibitem [{\citenamefont {Ginelli}\ \emph {et~al.}(2010)\citenamefont
  {Ginelli}, \citenamefont {Peruani}, \citenamefont {B\"ar},\ and\
  \citenamefont {Chat\'e}}]{Ginelli2012PRL}%
  \BibitemOpen
  \bibfield  {author} {\bibinfo {author} {\bibfnamefont {F.}~\bibnamefont
  {Ginelli}}, \bibinfo {author} {\bibfnamefont {F.}~\bibnamefont {Peruani}},
  \bibinfo {author} {\bibfnamefont {M.}~\bibnamefont {B\"ar}},\ and\ \bibinfo
  {author} {\bibfnamefont {H.}~\bibnamefont {Chat\'e}},\ }\bibfield  {title}
  {\bibinfo {title} {Large-scale collective properties of self-propelled
  rods},\ }\href {https://doi.org/10.1103/PhysRevLett.104.184502} {\bibfield
  {journal} {\bibinfo  {journal} {Phys. Rev. Lett.}\ }\textbf {\bibinfo
  {volume} {104}},\ \bibinfo {pages} {184502} (\bibinfo {year}
  {2010})}\BibitemShut {NoStop}%
\bibitem [{\citenamefont {Peruani}\ \emph {et~al.}(2012)\citenamefont
  {Peruani}, \citenamefont {Starru\ss{}}, \citenamefont {Jakovljevic},
  \citenamefont {S\o{}gaard-Andersen}, \citenamefont {Deutsch},\ and\
  \citenamefont {B\"ar}}]{Peruani2012PRL}%
  \BibitemOpen
  \bibfield  {author} {\bibinfo {author} {\bibfnamefont {F.}~\bibnamefont
  {Peruani}}, \bibinfo {author} {\bibfnamefont {J.}~\bibnamefont
  {Starru\ss{}}}, \bibinfo {author} {\bibfnamefont {V.}~\bibnamefont
  {Jakovljevic}}, \bibinfo {author} {\bibfnamefont {L.}~\bibnamefont
  {S\o{}gaard-Andersen}}, \bibinfo {author} {\bibfnamefont {A.}~\bibnamefont
  {Deutsch}},\ and\ \bibinfo {author} {\bibfnamefont {M.}~\bibnamefont
  {B\"ar}},\ }\bibfield  {title} {\bibinfo {title} {Collective motion and
  nonequilibrium cluster formation in colonies of gliding bacteria},\ }\href
  {https://doi.org/10.1103/PhysRevLett.108.098102} {\bibfield  {journal}
  {\bibinfo  {journal} {Phys. Rev. Lett.}\ }\textbf {\bibinfo {volume} {108}},\
  \bibinfo {pages} {098102} (\bibinfo {year} {2012})}\BibitemShut {NoStop}%
\bibitem [{\citenamefont {Ngo}\ \emph {et~al.}(2014)\citenamefont {Ngo},
  \citenamefont {Peshkov}, \citenamefont {Aranson}, \citenamefont {Bertin},
  \citenamefont {Ginelli},\ and\ \citenamefont {Chat\'e}}]{Nao2014PRL}%
  \BibitemOpen
  \bibfield  {author} {\bibinfo {author} {\bibfnamefont {S.}~\bibnamefont
  {Ngo}}, \bibinfo {author} {\bibfnamefont {A.}~\bibnamefont {Peshkov}},
  \bibinfo {author} {\bibfnamefont {I.~S.}\ \bibnamefont {Aranson}}, \bibinfo
  {author} {\bibfnamefont {E.}~\bibnamefont {Bertin}}, \bibinfo {author}
  {\bibfnamefont {F.}~\bibnamefont {Ginelli}},\ and\ \bibinfo {author}
  {\bibfnamefont {H.}~\bibnamefont {Chat\'e}},\ }\bibfield  {title} {\bibinfo
  {title} {Large-scale chaos and fluctuations in active nematics},\ }\href
  {https://doi.org/10.1103/PhysRevLett.113.038302} {\bibfield  {journal}
  {\bibinfo  {journal} {Phys. Rev. Lett.}\ }\textbf {\bibinfo {volume} {113}},\
  \bibinfo {pages} {038302} (\bibinfo {year} {2014})}\BibitemShut {NoStop}%
\bibitem [{\citenamefont {Nishiguchi}\ \emph {et~al.}(2017)\citenamefont
  {Nishiguchi}, \citenamefont {Nagai}, \citenamefont {Chat\'e},\ and\
  \citenamefont {Sano}}]{Nishiguchi2017PRE}%
  \BibitemOpen
  \bibfield  {author} {\bibinfo {author} {\bibfnamefont {D.}~\bibnamefont
  {Nishiguchi}}, \bibinfo {author} {\bibfnamefont {K.~H.}\ \bibnamefont
  {Nagai}}, \bibinfo {author} {\bibfnamefont {H.}~\bibnamefont {Chat\'e}},\
  and\ \bibinfo {author} {\bibfnamefont {M.}~\bibnamefont {Sano}},\ }\bibfield
  {title} {\bibinfo {title} {Long-range nematic order and anomalous
  fluctuations in suspensions of swimming filamentous bacteria},\ }\href
  {https://doi.org/10.1103/PhysRevE.95.020601} {\bibfield  {journal} {\bibinfo
  {journal} {Phys. Rev. E}\ }\textbf {\bibinfo {volume} {95}},\ \bibinfo
  {pages} {020601(R)} (\bibinfo {year} {2017})}\BibitemShut {NoStop}%
\bibitem [{\citenamefont {Kawaguchi}\ \emph {et~al.}(2017)\citenamefont
  {Kawaguchi}, \citenamefont {Kageyama},\ and\ \citenamefont
  {Sano}}]{Kawaguchi:2017}%
  \BibitemOpen
  \bibfield  {author} {\bibinfo {author} {\bibfnamefont {K.}~\bibnamefont
  {Kawaguchi}}, \bibinfo {author} {\bibfnamefont {R.}~\bibnamefont
  {Kageyama}},\ and\ \bibinfo {author} {\bibfnamefont {M.}~\bibnamefont
  {Sano}},\ }\bibfield  {title} {\bibinfo {title} {Topological defects control
  collective dynamics in neural progenitor cell cultures},\ }\href
  {https://doi.org/10.1038/nature22321} {\bibfield  {journal} {\bibinfo
  {journal} {Nature}\ }\textbf {\bibinfo {volume} {545}},\ \bibinfo {pages}
  {327} (\bibinfo {year} {2017})}\BibitemShut {NoStop}%
\bibitem [{\citenamefont {Mahault}\ \emph {et~al.}(2019)\citenamefont
  {Mahault}, \citenamefont {Ginelli},\ and\ \citenamefont
  {Chat\'e}}]{Mahault2019PRL}%
  \BibitemOpen
  \bibfield  {author} {\bibinfo {author} {\bibfnamefont {B.}~\bibnamefont
  {Mahault}}, \bibinfo {author} {\bibfnamefont {F.}~\bibnamefont {Ginelli}},\
  and\ \bibinfo {author} {\bibfnamefont {H.}~\bibnamefont {Chat\'e}},\
  }\bibfield  {title} {\bibinfo {title} {Quantitative assessment of the toner
  and tu theory of polar flocks},\ }\href
  {https://doi.org/10.1103/PhysRevLett.123.218001} {\bibfield  {journal}
  {\bibinfo  {journal} {Phys. Rev. Lett.}\ }\textbf {\bibinfo {volume} {123}},\
  \bibinfo {pages} {218001} (\bibinfo {year} {2019})}\BibitemShut {NoStop}%
\bibitem [{\citenamefont {Iwasawa}\ \emph {et~al.}(2021)\citenamefont
  {Iwasawa}, \citenamefont {Nishiguchi},\ and\ \citenamefont
  {Sano}}]{Iwasawa2021PRR}%
  \BibitemOpen
  \bibfield  {author} {\bibinfo {author} {\bibfnamefont {J.}~\bibnamefont
  {Iwasawa}}, \bibinfo {author} {\bibfnamefont {D.}~\bibnamefont
  {Nishiguchi}},\ and\ \bibinfo {author} {\bibfnamefont {M.}~\bibnamefont
  {Sano}},\ }\bibfield  {title} {\bibinfo {title} {Algebraic correlations and
  anomalous fluctuations in ordered flocks of janus particles fueled by an ac
  electric field},\ }\href {https://doi.org/10.1103/PhysRevResearch.3.043104}
  {\bibfield  {journal} {\bibinfo  {journal} {Phys. Rev. Research}\ }\textbf
  {\bibinfo {volume} {3}},\ \bibinfo {pages} {043104} (\bibinfo {year}
  {2021})}\BibitemShut {NoStop}%
\bibitem [{\citenamefont {Ginelli}(2016)}]{Ginelli:2016vx}%
  \BibitemOpen
  \bibfield  {author} {\bibinfo {author} {\bibfnamefont {F.}~\bibnamefont
  {Ginelli}},\ }\bibfield  {title} {\bibinfo {title} {The physics of the vicsek
  model},\ }\href {https://doi.org/10.1140/epjst/e2016-60066-8} {\bibfield
  {journal} {\bibinfo  {journal} {The European Physical Journal Special
  Topics}\ }\textbf {\bibinfo {volume} {225}},\ \bibinfo {pages} {2099}
  (\bibinfo {year} {2016})}\BibitemShut {NoStop}%
\bibitem [{\citenamefont {Dean}(1996)}]{Dean_1996}%
  \BibitemOpen
  \bibfield  {author} {\bibinfo {author} {\bibfnamefont {D.~S.}\ \bibnamefont
  {Dean}},\ }\bibfield  {title} {\bibinfo {title} {Langevin equation for the
  density of a system of interacting langevin processes},\ }\href
  {https://doi.org/10.1088/0305-4470/29/24/001} {\bibfield  {journal} {\bibinfo
   {journal} {Journal of Physics A: Mathematical and General}\ }\textbf
  {\bibinfo {volume} {29}},\ \bibinfo {pages} {L613} (\bibinfo {year}
  {1996})}\BibitemShut {NoStop}%
\bibitem [{\citenamefont {Nakamura}\ and\ \citenamefont
  {Yoshimori}(2009)}]{Nakamura_2009}%
  \BibitemOpen
  \bibfield  {author} {\bibinfo {author} {\bibfnamefont {T.}~\bibnamefont
  {Nakamura}}\ and\ \bibinfo {author} {\bibfnamefont {A.}~\bibnamefont
  {Yoshimori}},\ }\bibfield  {title} {\bibinfo {title} {Derivation of the
  nonlinear fluctuating hydrodynamic equation from the underdamped langevin
  equation},\ }\href {https://doi.org/10.1088/1751-8113/42/6/065001} {\bibfield
   {journal} {\bibinfo  {journal} {Journal of Physics A: Mathematical and
  Theoretical}\ }\textbf {\bibinfo {volume} {42}},\ \bibinfo {pages} {065001}
  (\bibinfo {year} {2009})}\BibitemShut {NoStop}%
\bibitem [{\citenamefont {Zwillinger}(2014)}]{zwillinger2014}%
  \BibitemOpen
  \bibfield  {author} {\bibinfo {author} {\bibfnamefont {D.}~\bibnamefont
  {Zwillinger}},\ }\href {https://books.google.co.jp/books?id=NjnLAwAAQBAJ}
  {\emph {\bibinfo {title} {Table of Integrals, Series, and Products}}}\
  (\bibinfo  {publisher} {Elsevier Science},\ \bibinfo {year}
  {2014})\BibitemShut {NoStop}%
\bibitem [{\citenamefont {Frisch}(1995)}]{frisch_1995}%
  \BibitemOpen
  \bibfield  {author} {\bibinfo {author} {\bibfnamefont {U.}~\bibnamefont
  {Frisch}},\ }\href {https://doi.org/10.1017/CBO9781139170666} {\emph
  {\bibinfo {title} {Turbulence: The Legacy of A. N. Kolmogorov}}}\ (\bibinfo
  {publisher} {Cambridge University Press},\ \bibinfo {year}
  {1995})\BibitemShut {NoStop}%
\bibitem [{\citenamefont {Caprini}\ and\ \citenamefont {Marini
  Bettolo~Marconi}(2020)}]{Caprini2020jcp}%
  \BibitemOpen
  \bibfield  {author} {\bibinfo {author} {\bibfnamefont {L.}~\bibnamefont
  {Caprini}}\ and\ \bibinfo {author} {\bibfnamefont {U.}~\bibnamefont {Marini
  Bettolo~Marconi}},\ }\bibfield  {title} {\bibinfo {title} {Active matter at
  high density: Velocity distribution and kinetic temperature},\ }\href
  {https://doi.org/10.1063/5.0029710} {\bibfield  {journal} {\bibinfo
  {journal} {The Journal of Chemical Physics}\ }\textbf {\bibinfo {volume}
  {153}},\ \bibinfo {pages} {184901} (\bibinfo {year} {2020})}\BibitemShut
  {NoStop}%
\bibitem [{\citenamefont {Gardiner}(2009)}]{gardiner}%
  \BibitemOpen
  \bibfield  {author} {\bibinfo {author} {\bibfnamefont {C.}~\bibnamefont
  {Gardiner}},\ }\href {https://books.google.co.jp/books?id=otg3PQAACAAJ}
  {\emph {\bibinfo {title} {Stochastic Methods: A Handbook for the Natural and
  Social Sciences}}}\ (\bibinfo  {publisher} {Springer Berlin Heidelberg},\
  \bibinfo {year} {2009})\BibitemShut {NoStop}%
\bibitem [{\citenamefont {Fodor}\ \emph {et~al.}(2016)\citenamefont {Fodor},
  \citenamefont {Nardini}, \citenamefont {Cates}, \citenamefont {Tailleur},
  \citenamefont {Visco},\ and\ \citenamefont {van Wijland}}]{Fodor2016PRL}%
  \BibitemOpen
  \bibfield  {author} {\bibinfo {author} {\bibfnamefont {E.}~\bibnamefont
  {Fodor}}, \bibinfo {author} {\bibfnamefont {C.}~\bibnamefont {Nardini}},
  \bibinfo {author} {\bibfnamefont {M.~E.}\ \bibnamefont {Cates}}, \bibinfo
  {author} {\bibfnamefont {J.}~\bibnamefont {Tailleur}}, \bibinfo {author}
  {\bibfnamefont {P.}~\bibnamefont {Visco}},\ and\ \bibinfo {author}
  {\bibfnamefont {F.}~\bibnamefont {van Wijland}},\ }\bibfield  {title}
  {\bibinfo {title} {How far from equilibrium is active matter?},\ }\href
  {https://doi.org/10.1103/PhysRevLett.117.038103} {\bibfield  {journal}
  {\bibinfo  {journal} {Phys. Rev. Lett.}\ }\textbf {\bibinfo {volume} {117}},\
  \bibinfo {pages} {038103} (\bibinfo {year} {2016})}\BibitemShut {NoStop}%
\end{thebibliography}
\end{document}